\documentclass{article}

\usepackage[preprint]{neurips_2026}

\usepackage{booktabs}
\usepackage{amsmath,amssymb}
\usepackage{algorithm}
\usepackage{algorithmic}
\usepackage{graphicx}
\usepackage[table]{xcolor}
\usepackage{multirow}
\usepackage{subcaption}
\usepackage{afterpage}
\usepackage{natbib}
\usepackage{enumitem}
\usepackage[utf8]{inputenc}
\usepackage[T1]{fontenc}
\usepackage{hyperref}
\usepackage{url}
\usepackage{cleveref}
\usepackage{makecell}
\usepackage{mdframed}

\newmdenv[
  linecolor=black,
  backgroundcolor=gray!10,
  roundcorner=5pt,
  innertopmargin=1ex,
  innerbottommargin=1ex
]{casebox}

\newcounter{customalg}

\newcommand{\stategraph}{\mathcal{G}}
\newcommand{\stateset}{\mathcal{S}}
\newcommand{\edgeset}{\mathcal{E}}
\newcommand{\noisegraph}{\hat{\mathcal{G}}}

\newcommand{\testcase}{T}
\newcommand{\instruction}{I}

\newcommand{\vcase}{V_{\text{case}}}
\newcommand{\traj}{\tau}

\newcommand{\Prob}{\mathbb{P}}
\usepackage[most]{tcolorbox}

\definecolor{PromptFrame}{RGB}{78,78,78}
\definecolor{PromptBody}{RGB}{245,245,245}
\definecolor{PromptTitle}{RGB}{72,72,72}
\definecolor{PromptRule}{RGB}{110,110,110}
\definecolor{PromptText}{HTML}{1F2937}

\newtcolorbox{promptbox}[2][]{%
  enhanced,
  breakable,
  colback=PromptBody,
  colframe=PromptFrame,
  boxrule=0.9pt,
  sharp corners,
  left=4mm,
  right=3mm,
  top=2.5mm,
  bottom=2.5mm,
  colbacktitle=PromptTitle,
  coltitle=white,
  title={#2},
  fonttitle=\bfseries\small,
  title filled,
  titlerule=0.45pt,
  titlerule style={PromptRule},
  lefttitle=4mm,
  righttitle=4mm,
  toptitle=1.4mm,
  bottomtitle=1.3mm,
  #1
}

\usepackage[normalem]{ulem}
\newif\ifshowcomments
\showcommentsfalse

\ifshowcomments
  \newcommand{\zhijie}[1]{\textcolor{blue}{[\textbf{ZJ:} #1]}}
  \newcommand{\yifan}[1]{\textcolor{brown}{[\textbf{WYF:} #1]}}
  \newcommand{\stella}[1]{\textcolor{magenta!65!black}{[\textbf{SR:} #1]}}
  \newcommand{\tengfei}[1]{\textcolor{teal}{[\textbf{TF:} #1]}}
  \newcommand{\todo}[1]{\textcolor{orange}{[\textbf{TODO:} #1]}}
  
  \newcommand{\del}[1]{\textcolor{gray}{\sout{#1}}}
\else
  \newcommand{\zhijie}[1]{}
  \newcommand{\yifan}[1]{}
  \newcommand{\stella}[1]{}
  \newcommand{\tengfei}[1]{}
  \newcommand{\todo}[1]{}
  
  \newcommand{\del}[1]{}
\fi

\title{DiagEval: Trajectory-Conditioned Diagnosis for Reliable Software Evaluation with GUI Agents}

\author{%
\small
Sirui Hong\textsuperscript{1}\thanks{Equal contribution.}
\quad
Zhijie Liu\textsuperscript{1}\footnotemark[1]
\quad
Tengfei Li\textsuperscript{1}
\quad
Wei Tao\textsuperscript{2}
\quad
Yifan Wu\textsuperscript{3}
\quad
Chenglin Wu\textsuperscript{1}\thanks{Corresponding author: \texttt{alexanderwu@deepwisdom.ai}.}
\\[-0.1em]
\footnotesize
\textsuperscript{1}DeepWisdom
\quad
\textsuperscript{2}Independent Researcher
\quad
\textsuperscript{3}The Hong Kong University of Science and Technology (Guangzhou)
\vspace{-0.4em}
}

\begin{document}

\maketitle

\begin{abstract}
Evaluating LLM-generated interactive software requires execution in addition to static analysis. The key difficulty is that correctness is a graph-level reachability property over latent UI state-transition graphs, whereas a GUI evaluator observes only a single execution trajectory. A failed rollout therefore rules out only one realized path, leaving failure attribution ambiguous between evaluator-side execution error and genuine software defect.
We present \textsc{DiagEval}, a trajectory-conditioned diagnostic evaluation protocol for post-failure GUI-agent evaluation of interactive software. Rather than blindly retrying from scratch, \textsc{DiagEval} reuses the failed trajectory to choose targeted diagnostic probes and aggregates their outcomes into an internal attribution signal. The latent-graph view motivates the diagnostic problem; \textsc{DiagEval} does not reconstruct the graph or estimate calibrated posterior probabilities.
We evaluate \textsc{DiagEval} on WebDevJudge-Unit and RealDevBench across multiple GUI-agent evaluators and LLM backbones. On false-negative cases, \textsc{DiagEval} recovers 45.6--62.1\% of failures that were initially misattributed to software defects, outperforming retry-based baselines with 34.4--160.6\% relative gains. On the full evaluation sets, this recovery improves accuracy from 69.9\% to 78.3\% on WebDevJudge-Unit and from 65.0\% to 81.6\% on RealDevBench. These results suggest that reliable GUI-agent evaluation requires not only stronger execution, but also active failure diagnosis to disambiguate evaluator-side errors from genuine software defects. Our code is available at \url{https://github.com/scutGit/DiagEval}.
\end{abstract}

\section{Introduction}
\label{sec:intro}
As LLM-generated software grows from isolated code snippets to full-stack applications with complex UIs, the evaluation problem changes with it: static analysis is no longer enough, and correctness must be verified through interactive execution~\citep{bian2025dontknowclickautomatedgui,lu2025webgenbench,xiao2025codeaesthetics,li2026webdevjudge,kong2026webtestbench,PlayCoder2026}. GUI agents are a natural choice for this role. However, they introduce a new reliability risk: evaluator-side failures can become false evidence against the software itself.
This is not a corner case. Recent works~\citep{li2026webdevjudge,lu2025webgenbench} on automated web-development evaluation show that agentic evaluators can incorrectly label feasible tasks as infeasible due to their own operational failures rather than actual shortcomings in the implementation. In parallel, exploratory GUI testing work identifies an execution-bias attribution problem, where software-side defects are misidentified as agent-side execution slips~\citep{gao2026guitesterenablingguiagents}.
This ambiguity is further amplified by agent-environment misalignment: as noted by \citet{liu2025agent}, an agent's expected outcome may diverge from the environment's actual state transition. For example, a button that remains unchanged after a click may indicate a broken interface element, a misgrounded action, or a view that failed to expose the relevant state transition.

At its core, the challenge is \emph{failure attribution under limited trajectory evidence}: a failed GUI-agent rollout records only one agent--environment interaction trace, in which actions, observations, DOM states, logs, and reasoning traces are entangled. These signals are diagnostically ambiguous because the same failed transition can arise from different sources: the agent may ground or execute an action incorrectly, the environment may expose only a partial observation of the relevant program state, or the software may genuinely block all valid progress.
This ambiguity is structural rather than incidental, because software correctness is a graph-level property while an evaluator observes only a path-local slice of it. 
A test should be considered successful if at least one valid execution path reaches the target state in the software's latent state-transition graph. By contrast, a failed rollout rules out only the sampled path and does not prove that the goal state is unreachable. 
We refer to this structural under-determination as the \emph{single-trajectory identifiability gap}.
After observing a single failed trajectory, the latent cause of failure is therefore structurally underdetermined. The failure may reflect evaluator-side under-exploration or execution error (\textsc{AgentFail}), or it may indicate a genuine system defect (\textsc{EnvFail}). Without probing alternative paths or selectively acquiring additional evidence, these hypotheses remain observationally confounded.

Existing agent frameworks and reliability mechanisms improve GUI-agent execution in different ways, but they do not directly address post-failure attribution.
ReAct-style agents~\citep{yao2023react} interleave reasoning and acting to gather evidence, but they do not explicitly distinguish execution unreliability from genuine software failure after a failed interaction. Repetition- and feedback-based strategies such as naive retry, Best-of-$N$ sampling, and self-correction can reduce transient errors by adding more attempts or feedback, but they often fail when the same agent-side mistake recurs across runs or is reinforced by correlated self-critique~\citep{shinn2023reflexion,madaan2023selfrefine,huang2024selfcorrect}. 
More broadly, robustness methods acknowledge environment noise and uncertainty, but aim to harden policies against it rather than attribute a particular failure~\citep{smirnova2019distributionallyrobustreinforcementlearning}. Post hoc attribution and active information acquisition offer related tools for diagnosing completed traces or selecting informative actions, but they are not designed to turn a failed GUI-agent rollout into online diagnostic probes for software evaluation~\citep{zhang2025agentracer,veiga2023reactive,chen2026seeing}.
What is missing is a diagnostic stance for GUI-agent evaluation: after a failure, the evaluator should use the failed trajectory to decide which uncertainty sources to probe next and how the resulting evidence should be attributed to evaluator-side error versus genuine software defect.

To address this, we propose \textsc{DiagEval}, which formulates post-failure software evaluation as active diagnosis under attribution uncertainty. Rather than retrying from scratch, \textsc{DiagEval} reuses the executed trajectory to identify the dominant source of uncertainty, interacts with the environment through targeted probes to disambiguate uncertainty, and integrates the resulting outcomes into an internal attribution signal. In this way, post-failure evaluation becomes a trajectory-conditioned diagnostic process rather than undifferentiated re-execution.
Across two interactive software benchmarks, \textsc{DiagEval} improves recovery and accuracy over retry-based baselines, and its diagnostic mechanism transfers across GUI-agent frameworks, supporting post-failure diagnosis as an evaluator-side reliability mechanism.
Our contributions are as follows:
\begin{enumerate}[nosep,leftmargin=*]
    \item \textbf{Problem formulation.} We identify post-failure attribution under limited trajectory evidence as a distinct reliability problem in interactive software evaluation. We show that single-trajectory verdicts are structurally insufficient because graph-level correctness cannot be determined from path-local observation alone.

    \item \textbf{Trajectory-conditioned diagnostic protocol.} We propose \textsc{DiagEval}, which decomposes a failed trajectory into typed sources of uncertainty, prioritizes candidate diagnostic probes via a structured information-value ranking, and integrates branch-typed outcomes into an internal attribution signal through an update rule.

    \item \textbf{Reliability gains across benchmarks.} On WebDevJudge-Unit (WDJ-U)~\citep{li2026webdevjudge} and RealDevBench (RDB)~\citep{bian2025dontknowclickautomatedgui}, \textsc{DiagEval} recovers 45.6--62.1\% of false negatives, compared with 17.5--46.2\% under retry-based baselines. On the full evaluation sets, accuracy rises from 69.9\% to 78.3\% on WDJ-U and from 65.0\% to 81.6\% on RDB.

    \item \textbf{Reliability--cost frontier and transfer.} \textsc{DiagEval} improves reliability without scaling the GUI-execution backbone or specializing to a GUI-agent framework. With \texttt{gemini-3-flash-preview} as the executing GUI agent, it reaches 76.1\%/78.6\% on WDJ-U/RDB, exceeding AppEvalPilot run end-to-end on \texttt{claude-opus-4-6} at less than half its per-case cost; it also transfers to UI-TARS without retuning, with $+14.2$--$+22.1$ accuracy points.
\end{enumerate}
\section{Related Work}
\label{sec:related}

\vspace{-0.5em}
\subsection{Evaluation for Software Engineering}
\vspace{-0.3em}
Software-engineering evaluation has progressed from function- and repository-level code benchmarks~\citep{chen2021humaneval,zhuo2024bigcodebench,jain2025livecodebench,zhang2024naturalcodebench,ding2023crosscodeeval,liu2024repobench,jimenez2024swebench,miserendino2025swe} to agentic, interactive settings~\citep{zhuge2024agent,chan2024mle,zhou2024webarena,xie2024osworld,rawles2024androidworld}. Most relevant to us, WebDevJudge and RealDevBench~\citep{li2026webdevjudge,bian2025dontknowclickautomatedgui} adopt execution-based evaluation in dynamic environments, yet still treat a failed rollout as a terminal verdict. We instead study post-failure attribution: whether a failure reflects a true software defect or evaluator-side under-exploration.

\vspace{-0.8em}
\subsection{LLM-as-a-Judge}
\vspace{-0.3em}
LLM-as-a-Judge~\citep{NEURIPS2023_91f18a12} evaluates static outputs, and Agent-as-a-Judge~\citep{zhuge2024agent} extends this paradigm to intermediate trajectories. However, both treat evaluation as scoring a fixed artifact. Existing efforts to improve reliability, e.g., debiasing, hallucination reduction, judge finetuning, rubric prompting, and multi-judge ensembling~\citep{ye2024justiceprejudicequantifyingbiases, li2025llmsreliablyjudgeyet, chen2024mllmasajudge, thakur-etal-2025-judging, xu-etal-2023-instructscore, wei2025rocketevalefficientautomatedllm, chan2023chatevalbetterllmbasedevaluators, zhu2025judgelmfinetunedlargelanguage}, still assume fixed evidence. This assumption is inadequate for interactive software evaluation, where a failed trajectory exposes only a path-local slice of graph-level correctness. Recent Computer-Using Agent verification methods also improve judgment within the fixed-trajectory setting~\citep{rosset2026art}. 
By contrast, we treat failed rollouts as opportunities for active evidence acquisition, using targeted diagnostic probes to resolve agent-vs-environment attribution.

\vspace{-0.8em}
\subsection{GUI Agents}
\vspace{-0.3em}
GUI agents have progressed from metadata-based systems to fully visual agents that can autonomously interact with software environments~\citep{cheng2024seeclick,wu2024atlas,xu2024aguvis,gou2024navigating}, enabled by multimodal LLMs such as Claude 3.5 Sonnet~\citep{anthropic2024claude35sonnet} and Gemini 2.5~\citep{google2025gemini25cum}. However, as evaluators, GUI agents remain unreliable under partial observability, asynchronous rendering, and agent-induced execution errors, including hallucination, UI mislocalization, and inconsistent behavior in dynamic environments~\citep{jin2026halluclear,li2026webdevjudge}. Frameworks such as WebGen-Bench~\citep{lu2025webgenbench} and UXAgent~\citep{lu2025uxagent} exploit GUI agents for large-scale interactive evaluation, but their judgments are still trajectory-dependent and lack attribution capability: they cannot reliably distinguish software defects from evaluator-side failures. 
We address this limitation by formulating evaluator failure as a trajectory-level diagnosis problem and using cross-trajectory evidence to distinguish software defects from agent-side execution misses.
\vspace{-0.5em}
\section{Preliminaries and Problem Formulation}
\label{sec:prelim}
\vspace{-0.45em}

\subsection{Graph-Based Evaluation as Reachability Inference}
\label{sec:graph}
\begin{figure}[!t]
    \centering
    \includegraphics[width=0.95\textwidth]{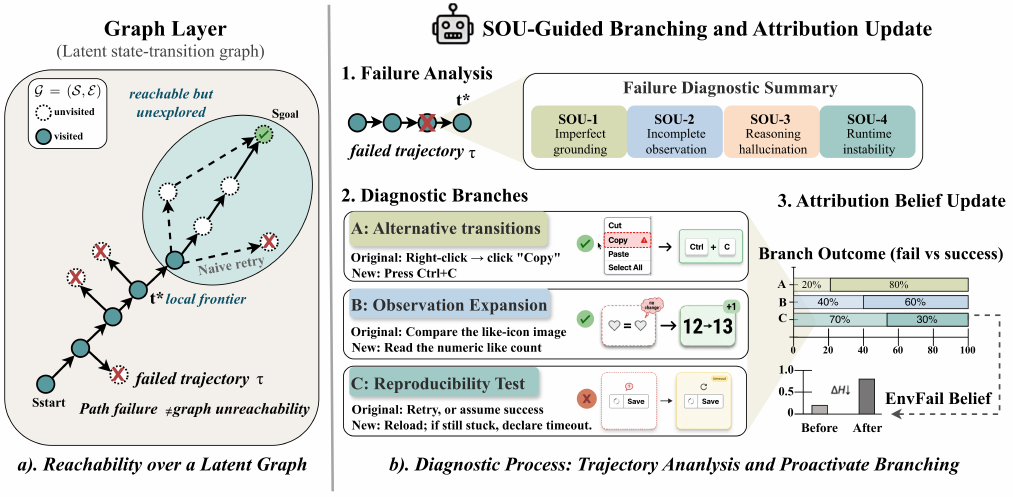}
    \caption{\textbf{Overview of \textsc{DiagEval}.}
Given a failed rollout $\traj$, \textsc{DiagEval} parses a failure diagnostic summary (FDS), dispatches SOU-guided diagnostic branches, and integrates evidence across multiple branch trajectories to refine an internal attribution score over $Z\in\{\textsc{AgentFail},\textsc{EnvFail}\}$.}
    \label{fig:diageval_overview}
    \vspace{-0.7em}
\end{figure}

\vspace{-0.35em}
We model the software under test as a latent state-transition graph $\stategraph = (\stateset,\edgeset)$, where each edge represents an action-labeled executable UI transition between latent interface states.
A test case $\testcase=(\instruction,O)$ is successful if some reachable state satisfies the target output condition. We use $\vcase \in \{0,1\}$ to denote this ground-truth success label. Software correctness is therefore a graph-level reachability property.
The evaluator, however, does not observe $\stategraph$ directly. Instead, it interacts with the software through a partially observed trajectory $\traj=(o_0,a_0,\ldots,o_N)$, where each $o_t$ is an observation and each $a_t$ is an executed action. Based on this trajectory, the evaluator outputs a binary verdict $\hat{y}\in\{0,1\}$ indicating whether the test case is judged successful. The evaluation objective is therefore to maximize $\Prob(\hat{y}=\vcase)$.
The key mismatch is structural: the evaluator observes only the realized execution path, whereas correctness depends on whether some valid path exists in the underlying reachable graph.
As shown in Figure~\ref{fig:diageval_overview}(a), a failed trajectory $\traj$ rules out only the realized path, not the reachability of $s_{\mathrm{goal}}$.
As a result, a failed trajectory does not establish that the target state is unreachable, making negative verdicts particularly difficult to interpret in GUI-based software evaluation.

\vspace{-0.65em}
\subsection{Sources of Uncertainty and the Attribution Problem}
\label{sec:motivation}
\vspace{-0.35em}
When a GUI agent fails to complete a task, the latent failure cause $Z \in \{\textsc{AgentFail}, \textsc{EnvFail}\}$ is not directly observable. \textsc{AgentFail} indicates that $\vcase=1$, i.e., a valid execution path exists in $\stategraph$, but the evaluator fails to discover or verify it. \textsc{EnvFail} indicates that $\vcase=0$, i.e., the target state is genuinely unreachable under the intended task semantics. The evaluator must infer $Z$ from indirect evidence carried by the executed trajectory. Each rollout reveals only a noisy, partial local view of the latent transition graph, shaped by perception, reasoning, and runtime distortions.
We identify four systematic \emph{sources of uncertainty} (SOU) that challenge attribution:

\vspace{-0.2em}
\par\noindent\textbf{(SOU 1) Imperfect grounding (perception errors).} The agent may fail to correctly locate or interact with the intended UI elements~\citep{gou2024navigating, lee2025reguidedataefficientgui, chen-etal-2025-guicourse, yang-etal-2025-aria} due to visual ambiguity, dynamic layouts, or subtle affordance changes. Such errors can make an evaluator-side failure appear externally similar to an environment-side blockage.

\vspace{-0.2em}
\par\noindent\textbf{(SOU 2) Incomplete observation (partial observability).}
The agent's view captures only the currently visible UI state,
missing hidden elements such as collapsed menus, off-screen content,
or latent interface regions~\citep{11186079, garousi2025aipoweredsoftwaretestingtools}.
As a result, the evaluator may conclude failure even though valid interactions remain undiscovered.

\vspace{-0.2em}
\par\noindent\textbf{(SOU 3) Reasoning hallucination (inference errors).} LLM-based evaluators may prematurely conclude task completion, over-interpret weak UI signals, or rationalize failed interactions as environmental faults~\citep{anthropic2025harnesses, zhang2025miragebenchllmagenthallucinating, chen2025trainingllmbasedagentssynthetic, lu2025webgenagentenhancinginteractivewebsite, ye2025aiagentswebtesting}. This introduces cognitive distortion on top of already partial evidence.

\vspace{-0.6em}
\par\noindent\textbf{(SOU 4) Runtime instability (execution errors).} Transient environment conditions such as network timeouts, nondeterministic backend latency, or flaky UI state may cause non-reproducible
failures, affecting action validity and outcome stability.

These heterogeneous, context-dependent SOUs cannot be reliably disambiguated from a single trajectory alone. Different SOUs imply different diagnostic responses: SOU-1/2/3 require testing alternative paths, while SOU-4 needs reproducibility checks. This makes attribution inherently diagnosis-dependent.

\vspace{-0.55em}
\subsection{The Single-Trajectory Identifiability Gap}
\label{sec:identifiability}
\vspace{-0.35em}
This ambiguity is structural. Given a failed trajectory $\traj$, the post-failure evaluation problem is to maintain evidence about $Z \in \{\textsc{AgentFail}, \textsc{EnvFail}\}$ from partial trajectory observations. This is particularly challenging for negative verdicts, since a single observed trajectory reveals only one realized path, while failure may arise either from an environment-side defect or from the evaluator's inability to discover an alternative feasible path. We refer to this underdetermination as the \emph{single-trajectory identifiability gap}, which motivates post-failure evaluation as a sequential diagnosis problem requiring active evidence acquisition.
\section{Method: DiagEval}
\label{sec:method}
\vspace{-0.5em}
We formulate post-failure software evaluation as a \emph{sequential diagnosis problem}: after an initial GUI-agent rollout receives a negative verdict $\hat{y}_1=0$, the evaluator gathers evidence to infer whether the failure reflects \textsc{AgentFail} or \textsc{EnvFail}. As illustrated in Figure~\ref{fig:diageval_overview}(b), \textsc{DiagEval} implements this diagnosis loop by parsing the failed trajectory into a Failure Diagnostic Summary (FDS), generating diagnostic branches from selected restart states, and then executing and aggregating branch evidence into an internal attribution score. This score guides subsequent diagnosis, but it is not a calibrated posterior estimate. \textsc{DiagEval} uses it without reconstructing the latent transition graph. The complete algorithmic procedure is given in Appendix~\ref{sec:appendix_algorithm}.

\vspace{-0.5em}
\subsection{Failure Parsing: Fork-Node Localization and State Abstraction}
\label{sec:diag_init}
\par\noindent\textbf{Fork-Node Localization.}
Given an initial failed trajectory $\traj_1=(o_0,a_0,\ldots,o_T)$, \textsc{DiagEval} seeks a \emph{fork node} $t^*$, the restart point from which targeted re-exploration is expected to most effectively reduce attribution uncertainty over $Z$. Intuitively, the fork node is not simply the most anomalous step, but the earliest actionable point at which alternative continuations may still separate \textsc{AgentFail} from \textsc{EnvFail}.
Conceptually, a useful fork node balances \emph{goal proximity} to a feasible continuation and \emph{attribution informativeness} for reducing uncertainty over $Z$:
\begin{equation}
    t^* \in \arg\max_{t \in \mathcal{T}_{\mathrm{cand}}}\ \widehat{\mathrm{Prog}}(t)\cdot \widehat{\Delta H}(t).
    \label{eq:fork_selection}
\end{equation}
where $\mathcal{T}_{\mathrm{cand}}\subseteq\{0,\ldots,T-1\}$ denotes the candidate restart steps in $\traj_1$, $\widehat{\mathrm{Prog}}(t)\in[0,1]$ measures estimated proximity to a plausible continuation, and $\widehat{\Delta H}(t)\geq 0$ measures expected reduction in attribution uncertainty.
Because $\stategraph$ and the true attribution distribution over $Z$ are not observable from a single failed trajectory, \textsc{DiagEval} treats Eq.~\ref{eq:fork_selection} as an idealized selection criterion rather than a closed-form objective. It approximates this criterion with a reflective LLM-based judge $\mathcal{J}$ using a compressed diagnostic summary $\mathcal{R}_1$ derived from the failed GUI-agent trajectory $\traj_1$. The summary aggregates the executed trajectory context, agent action explanations, and final failure evidence such as screenshots, DOM state, and judge output.
The judge $\mathcal{J}$ analyzes $\mathcal{R}_1$ and predicts $(t^*, h^*, \delta^*) \leftarrow \mathcal{J}(\mathcal{R}_1, \texttt{task})$,
where $t^*$ is the selected restart index, $h^*$ is the dominant SOU, and $\delta^*$ is a diagnostic explanation. In effect, $\mathcal{J}$ implicitly performs both candidate ranking and step selection over $\mathcal{T}_{\mathrm{cand}}$ from trajectory-level evidence.

\par\noindent\textbf{Failure Diagnostic Summary (FDS).}
Based on the fork-node reasoning, \textsc{DiagEval} constructs a Failure Diagnostic Summary (FDS): $\mathcal{F}_0 = (t^*, h^*, D_{t^*}, \mathcal{C}_{t^*})$,
where $D_{t^*} = (\delta^*, \mathcal{A}_{\text{cand}})$ provides the failure explanation and candidate policies, and $\mathcal{C}_{t^*}$ denotes the recoverable environment context associated with $t^*$, used for hot resumption without replaying the prefix $\traj_{1:t^*-1}$.
The FDS serves as the structured interface between failure parsing and downstream diagnostic probing. Implementation details are provided in Appendix~\ref{sec:appendix_fds_prompt}.

\vspace{-0.4em}
\subsection{Diagnostic Probe Generation}
\label{sec:probe_gen}
\begin{figure}[H]
    \centering
    \includegraphics[width=0.95\textwidth]{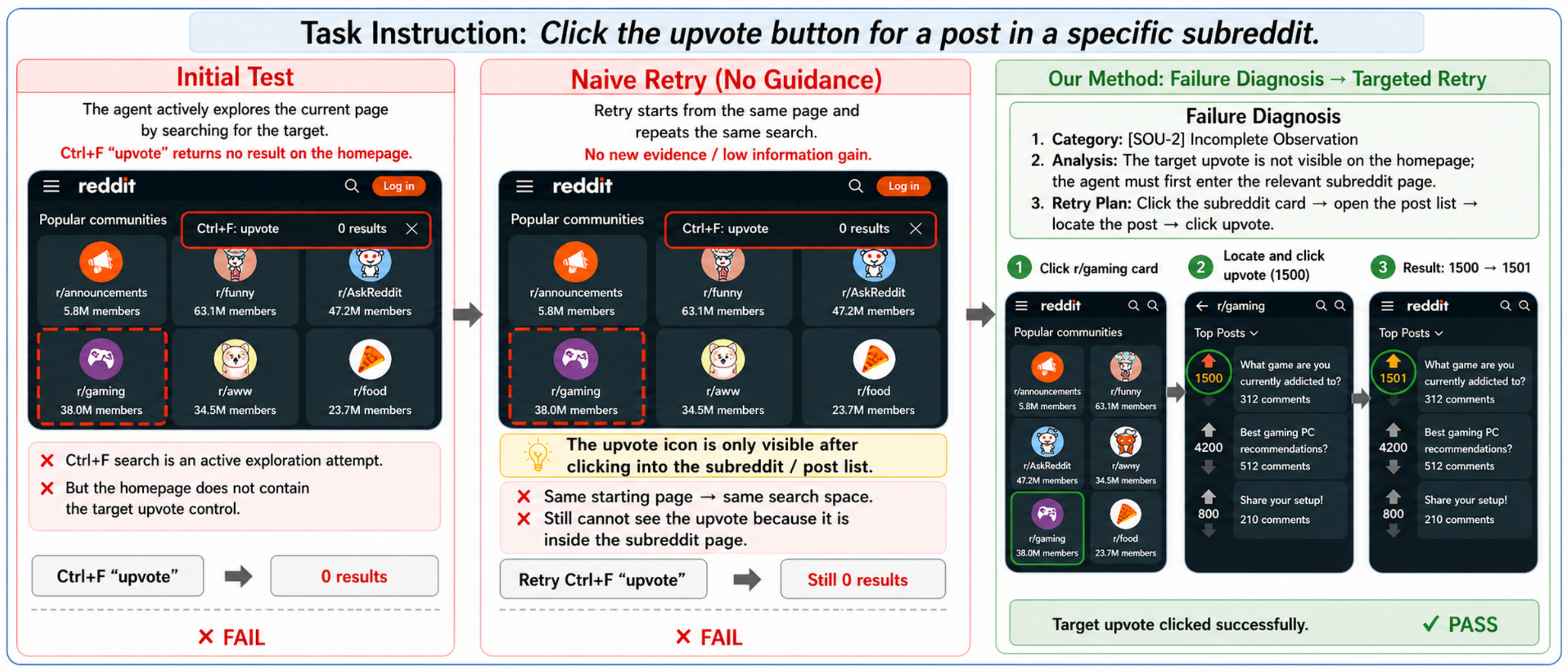}
    \caption{\textbf{Comparison of post-failure retry strategies} on \emph{``Click the upvote button for a post in a specific subreddit.''} The initial test (left) and a naive retry (middle) both fail along essentially the same homepage trajectory. \textsc{DiagEval} (right) uses the FDS to identify \textsc{SOU-2} (\emph{Incomplete Observation}), generates a targeted retry plan, and verifies success (\texttt{1500} $\to$ \texttt{1501}).}
    \label{fig:true_trajectory}
\vspace{-0.7em}
\end{figure}

\vspace{-0.4em}
Given the diagnostic state $\mathcal{F}_k=(t_k^*, h_k, D_{t_k^*}, \mathcal{C}_{t_k^*})$, \textsc{DiagEval} initializes probe types at the fork node $t_k^*$ and instantiates them as a candidate branch set $\mathcal{P}_k$. Here, a \emph{branch} denotes a detailed executable plan instantiated from a probe type. The dominant SOU $h_k$ guides this routing; details are provided in Appendices~\ref{sec:appendix_category_prompt} and~\ref{sec:appendix_branch_prompt}.
Figure~\ref{fig:true_trajectory} illustrates why diagnostic branches should be guided by the FDS rather than by blind retry: an unguided retry may reproduce the same visible state, whereas an FDS-guided branch redirects exploration toward more informative transitions.

\par\noindent\textbf{Type A (Alternative transitions).} Type-A probes realize the same subgoal through alternative actions, modalities, or routes. They test whether the original failure was caused by misexecution rather than genuine unreachability, and are primarily used when $h_k \in \{1,3\}$.
\vspace{-0.4em}
\par\noindent\textbf{Type B (Observation Expansion).} Type-B probes expand the visible frontier of the interface through scrolling, navigation, menu expansion, or state revelation. They test whether the original failure stemmed from incomplete observation, and are primarily used when $h_k = 2$.
\vspace{-0.4em}
\par\noindent\textbf{Type C (Reproducibility Test).} Type-C probes perform controlled repetitions under the same or minimally perturbed conditions to test whether the failure is stable. They are used when $h_k \in \{3,4\}$, especially for runtime instability and ambiguous reasoning failures.

Each branch $b \in \mathcal{P}_k$ contains: (i) a concrete action sequence, and (ii) an expected observation pattern that would differentially support \textsc{AgentFail} or \textsc{EnvFail}. These probe types are not only behaviorally distinct, but also attributionally asymmetric, which we formalize as branch-typed likelihoods in Sec.~\ref{sec:eig_ranking}.

\vspace{-0.85em}
\subsection{Information-Gain Branch Ranking}
\label{sec:eig_ranking}
\vspace{-0.4em}
Before execution in round $k$, \textsc{DiagEval} orders candidates in $\mathcal{P}_k$ by a structured information-value principle: branches are preferred when their possible outcomes are expected to better separate \textsc{AgentFail} from \textsc{EnvFail}. We instantiate this principle with a heuristic \textbf{Expected Information Gain} (EIG) score based on the branch-typed update rule in Sec.~\ref{sec:seq_update}.
Here, $p^{(k)}$ denotes \textsc{DiagEval}'s current preference toward \textsc{EnvFail} at diagnosis round $k$. To simulate outcomes, \textsc{DiagEval} uses likelihood parameters rather than calibrated probabilities. For a branch $b$ of type $d(b)$, $w_b$ scores recovery under \textsc{AgentFail}, $\beta_{d(b)}$ scores success that can still occur under \textsc{EnvFail}, and $\gamma_{d(b)}$ scores failure compatibility under \textsc{AgentFail}. We set $\gamma_A>\gamma_B>\gamma_C$, making Type~C failures stronger evidence for \textsc{EnvFail}. Formal parameter definitions are provided in Appendix~\ref{sec:appendix_selection_prompt}.
The EIG score of branch $b$ is then
\begin{equation}
\mathrm{EIG}(b;\mathcal{F}_k)
=
H_{\mathrm{bin}}(p^{(k)})
-
\mathbb{E}_{o\sim P(\cdot\mid b)}
\left[
H_{\mathrm{bin}}(p_{b,o}^{(k+1)})
\right],
\label{eq:eig}
\end{equation}
where $H_{\mathrm{bin}}(p^{(k)})$ is the attribution uncertainty before executing any branch in round $k$, and $p_{b,o}^{(k+1)}$ is the hypothetical attribution signal obtained if branch $b$ produced outcome $o$. The expectation is over $o\in\{\texttt{verified\_success},\texttt{fail}\}$ under the outcome scores defined above.
Consequently, branches expected to produce the largest reduction in attribution uncertainty are ranked first, while lower-gain branches are deferred, making the limited branch budget focus on more informative probes.

\subsection{Attribution Score Update and Stopping}
\label{sec:seq_update}
\vspace{-0.4em}
After EIG ranks the candidate set $\mathcal{P}_k$, \textsc{DiagEval} executes branches in descending EIG order. Let $p^{(k,j-1)}$ and $p^{(k,j)}$ denote the attribution signal before and after executing the $j$-th branch in round $k$. For branch $b_{k,j}$, let $d_{k,j}=d(b_{k,j})$, $w_{k,j}=w_{b_{k,j}}$, and $o_{k,j}$ be its observed outcome. The update rule follows Bayes' rule as a transparent scoring mechanism, not as a claim of calibrated probabilistic inference:
\begin{equation}
p^{(k,j)}=
\begin{cases}
\displaystyle\frac{p^{(k,j-1)}}{p^{(k,j-1)}+(1-p^{(k,j-1)})\gamma_{d_{k,j}}}, & o_{k,j}=\texttt{fail},\\[10pt]
\displaystyle\frac{\beta_{d_{k,j}}\, p^{(k,j-1)}}{w_{k,j}(1-p^{(k,j-1)})+\beta_{d_{k,j}}\, p^{(k,j-1)}}, & o_{k,j}=\texttt{verified\_success}.
\end{cases}
\label{eq:update_rule}
\end{equation}

The stopping rule reflects outcome asymmetry: \texttt{verified\_success} immediately stops diagnosis with verdict \textsc{Pass} ($\textsc{AgentFail}$) as a reachability witness. Failures are accumulated, and if the updated signal satisfies $p^{(k,j)} \geq \tau_{\mathrm{env}}$, \textsc{DiagEval} stops early with verdict \textsc{Fail} ($\textsc{EnvFail}$); otherwise, diagnosis continues with the updated attribution score until budget exhaustion.
\vspace{-0.5em}
\section{Experiments}
\label{sec:experiments}
\vspace{-0.5em}

\subsection{Experimental Setup}
\label{sec:setup}
\vspace{-0.1em}
\par\noindent\textbf{Datasets and Metrics.} We evaluate on two open-source web-development benchmarks: RealDevBench (RDB)~\citep{bian2025dontknowclickautomatedgui} with 429 annotated executable cases and WebDevJudge-Unit (WDJ-U)~\citep{li2026webdevjudge} with 502 unit-level tasks with ground-truth pass/fail labels.
We report accuracy and average cost per case (USD). On false-negative (FN) subsets, defined as cases with ground-truth \textsc{Pass} but initial verdict \textsc{Fail}, we additionally report FN recovery rate. Appendix~\ref{sec:appendix_dataset_scope} provides dataset details, and Appendix~\ref{sec:appendix_annotation} describes the RDB annotation protocol.

\par\noindent\textbf{Baselines.} We compare AppEvalPilot~\citep{bian2025dontknowclickautomatedgui}, WebVoyager~\citep{he2024webvoyager}, UI-TARS Agent~\citep{qin2025uitars}\footnote{We use the UI-TARS Agent implementation released with the WebDevJudge open-source repository.}, and model-native CUA. We report model-native CUA for Qwen3.5-35B-A3B, Claude Opus 4.6, and Gemini 3 Flash Preview~\citep{qwen2026qwen35,anthropic2026claudeopus46,google2025gemini3flash}, and evaluate GUI-agent frameworks with Claude Opus 4.6 or Gemini 3 Flash Preview as shown in Table~\ref{tab:overall_summary}. We set the GUI-agent step budget \emph{max\_iters} = 15 and 20 on WDJ-U and RDB, respectively.

\par\noindent\textbf{Diagnostic Settings.}
We use AppEvalPilot as the GUI-agent evaluation framework, invoking \textsc{DiagEval}'s online diagnosis procedure from Sec.~\ref{sec:method} only when an initial \textsc{Fail} verdict is returned.
We evaluate two diagnostic budgets: \textsc{DiagEval}\,($\times 1$) runs one diagnosis round, and \textsc{DiagEval}\,($\times 2$) adds a second round for cases that remain unresolved after the first.
Each round $k$ instantiates a candidate pool $\mathcal{P}_k$ of $N=5$ branches and executes the top $K=3$ under EIG ranking; diagnosis starts from a neutral prior $p^{(0)}=0.5$ and terminates upon \texttt{verified\_success} or once $p^{(k,j)}\geq\tau_{\mathrm{env}}=0.7$.
We set $(w_0,\beta_0)=(0.60,0.20)$ and $(\gamma_A,\gamma_B,\gamma_C)=(0.60,0.50,0.40)$, and instantiate the diagnostic judge $\mathcal{J}$ with Claude Sonnet 4.6~\citep{anthropic2026claudesonnet46}.
Per-case costs include the AppEvalPilot rollout and $\mathcal{J}$ calls for FDS construction and branch selection.
Appendix~\ref{app:model_call_composition} provides an auxiliary analysis of LLM-call composition on the FN diagnostic subset.
The parameter-setting protocol and sensitivity analyses are reported in Appendix~\ref{app:ablation_sensitivity}.

\vspace{-0.65em}
\subsection{Experimental Results}
\label{sec:main}

\vspace{-0.1em}
\par\noindent\textbf{(i) Reliability through targeted recovery.}
Table~\ref{tab:overall_summary} reports the main end-to-end comparison. With Gemini 3 Flash Preview as the GUI-execution backbone, \textsc{DiagEval} ($\times 1$) reaches 76.1\% accuracy on WDJ-U and 78.6\% on RDB, compared with 69.9\% and 65.0\% for AppEvalPilot. A second diagnostic round raises the accuracy to 78.3\% and 81.6\%, respectively. The accuracy gain is concentrated in false-negative recovery. After one round, FNs decrease by 45.6\% on WDJ-U ($114\rightarrow62$) and 47.0\% on RDB ($132\rightarrow70$), while 89.2\% and 94.5\% of the original TNs are preserved. After two rounds, only 40.4\% and 37.9\% of the original FNs remain, and TN preservation is still 86.0\% on WDJ-U and 84.9\% on RDB. This pattern indicates that \textsc{DiagEval} separates recoverable \textsc{AgentFail} cases from confirmed \textsc{EnvFail} cases among initially negative verdicts.

\vspace{-0.2em}
\par\noindent\textbf{(ii) Accuracy--cost tradeoff.}
On WDJ-U, \textsc{DiagEval} ($\times 1$) with Gemini 3 Flash Preview reaches 76.1\% accuracy at \$0.122 per case, exceeding AppEvalPilot with Claude Opus 4.6 at 30.5\% of its cost. The RDB comparison is more pronounced: the same diagnostic setting reaches 78.6\%, compared with 69.9\% for AppEvalPilot with Claude Opus 4.6, using only 47.5\% of the cost. These results suggest that post-failure attribution improves reliability beyond the gains obtainable from stronger GUI-agent execution alone.

\vspace{-0.1em}
\par
\begin{figure}[!b]
  \centering
  \begin{subfigure}[t]{0.5\textwidth}
      \centering
      \includegraphics[width=\linewidth]{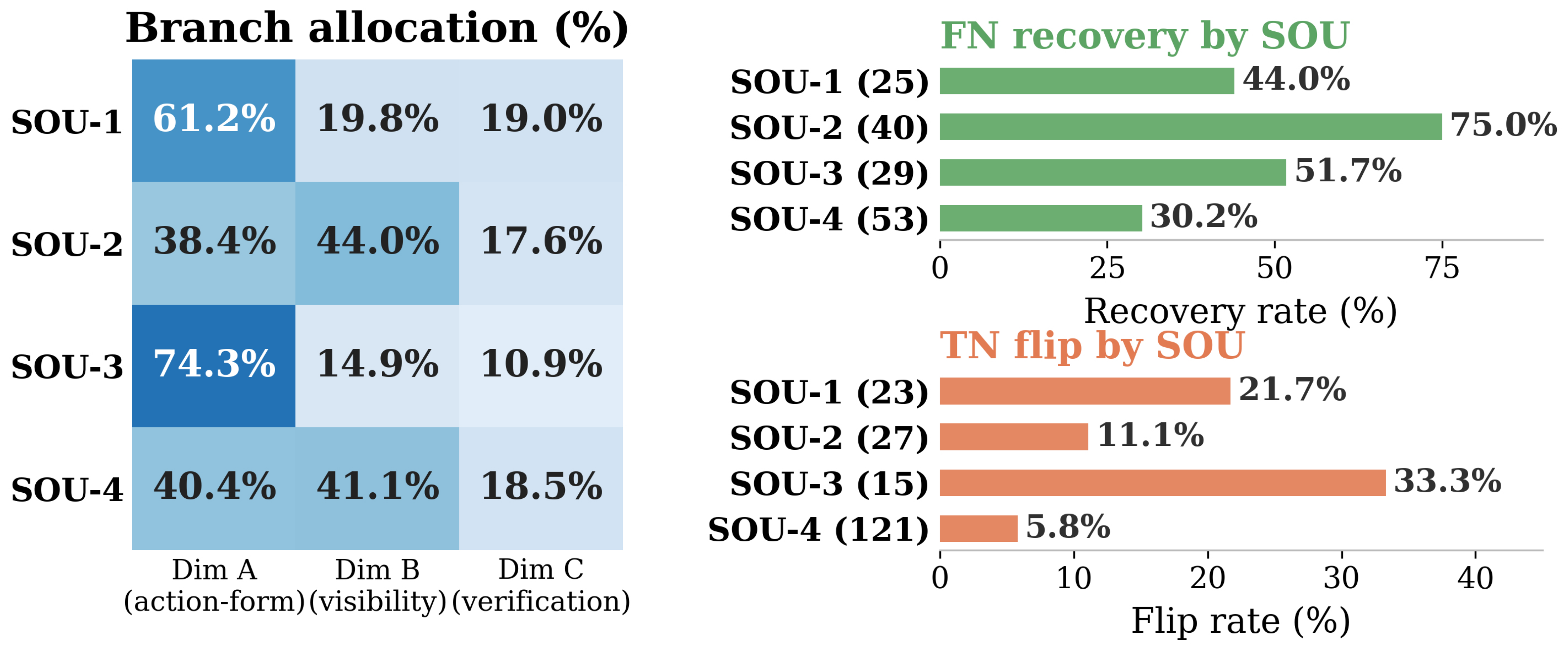}
      \caption{WDJ-U.}
      \label{fig:sou_branch_mechanism_wdj}
  \end{subfigure}\hfill
  \begin{subfigure}[t]{0.5\textwidth}
      \centering
      \includegraphics[width=\linewidth]{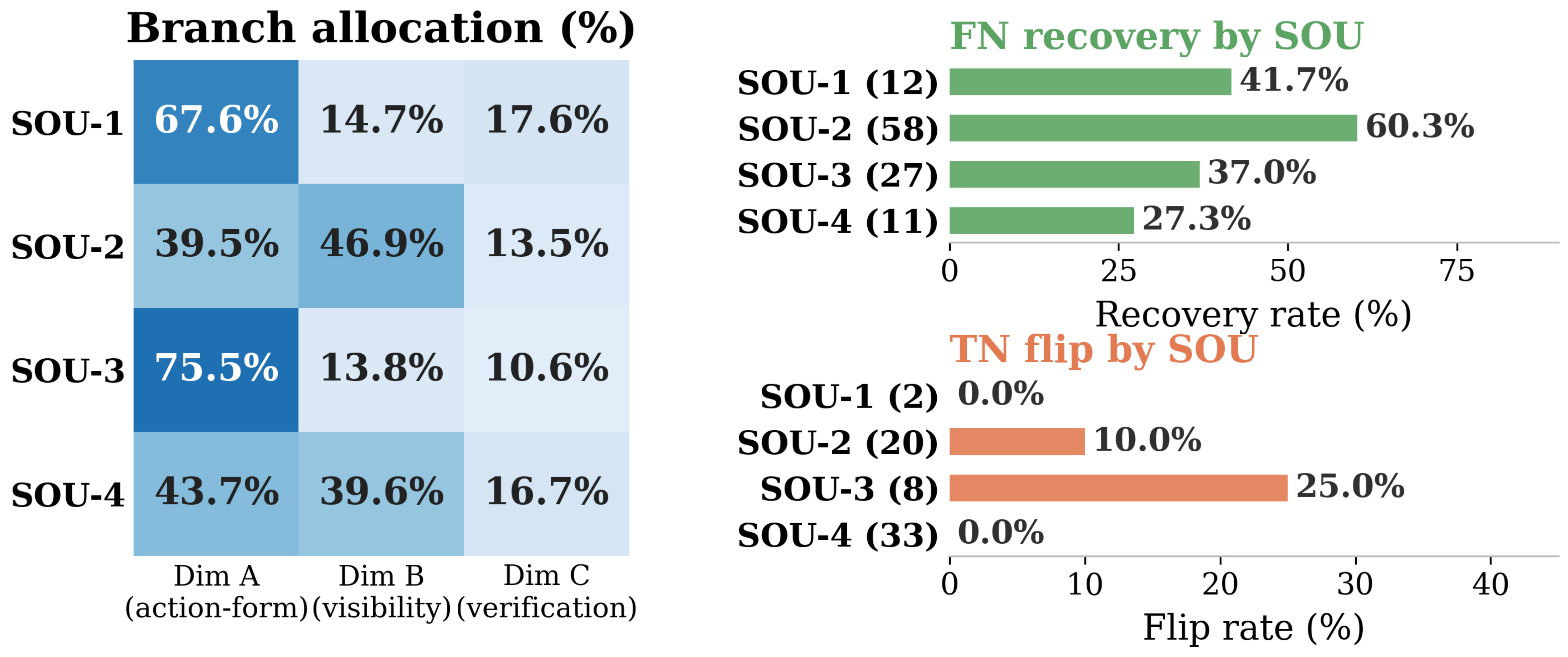}
      \caption{RDB.}
      \label{fig:sou_branch_mechanism_rdb}
  \end{subfigure}
  \caption{\textbf{SOU-typed branch allocation and diagnostic outcomes.}
Each subfigure corresponds to one benchmark. The left panel shows the distribution of Type~A/B/C probes within each SOU, and the right panels report FN recovery for $Z{=}\textsc{AgentFail}$ and TN flip for confirmed $Z{=}\textsc{EnvFail}$ cases.}
  \label{fig:sou_branch_mechanism}
\vspace{-0.8em}
\end{figure}

\noindent\textbf{(iii) Attribution evidence from SOU-guided probes.}
Figure~\ref{fig:sou_branch_mechanism} breaks down where the recovered cases come from. SOU-2 contributes the most to FN recovery, mainly through Type~B visibility branches, and achieves the highest recovery on both WDJ-U and RDB (75.0\% and 60.3\%). SOU-1 and SOU-3 mainly trigger Type~A alternative-transition branches, suggesting evaluator-side grounding or plan-realization errors. SOU-4 behaves differently: reproducibility-oriented branches produce a TN flip of $0.0\%$ on RDB and $5.8\%$ on WDJ-U within the SOU-4 subset. Overall, the accuracy gain comes from structured diagnostic evidence and informative counter-evidence, rather than from indiscriminately flipping failed verdicts.

\begin{table*}[t]
  \centering
  \caption{Main results on WDJ-U and RDB.
  \texttt{Acc.} is end-to-end accuracy, \texttt{$\Delta$} is the absolute change relative to AppEvalPilot, and FN/TN report false-negative and true-negative counts.
  Costs are reported in USD per case. $*$ indicates that the AppEvalPilot-relative change is not computed.
  Arrows indicate metric direction; best values are bolded, and \textsc{DiagEval} rows are highlighted.}
  \label{tab:overall_summary}
  \scriptsize
  \setlength{\tabcolsep}{2.2pt}
  \renewcommand{\arraystretch}{0.78}
  \begin{tabular*}{\textwidth}{@{\extracolsep{\fill}}p{3.45cm}p{3.15cm}>{\raggedleft\arraybackslash}p{1.45cm}>{\raggedleft\arraybackslash}p{1.35cm}rr>{\raggedleft\arraybackslash}p{1.2cm}@{}}
  \toprule
  \textbf{Model} & \textbf{Framework} & \textbf{Acc.}~$\uparrow$ & \textbf{$\Delta$}~$\uparrow$ & \textbf{FN}~$\downarrow$ & \textbf{TN}~$\uparrow$ & \textbf{\$/case}~$\downarrow$ \\
  \midrule
  \rowcolor{black!8}
  \multicolumn{7}{c}{\emph{(a) WDJ-U (502 cases)}} \\
  \midrule
  Qwen3.5-35B-A3B
    & Model CUA                                              & 46.1\%                  & *         & 258          & 211           & 0.08            \\
  \cmidrule(lr){1-7}
  \multirow{4}{*}{Claude Opus 4.6}
    & Model CUA                                              & 70.9\%                     & -3.6         & 88         & 174          & 0.32            \\
    & AppEvalPilot                                           & 74.1\%                  & ---         & 108         & \textbf{199} & 0.400          \\
    & UI-TARS                                                & 74.5\%                  & +0.4        & 102         & 198          & 0.361            \\
    & WebVoyager                                             & 65.5\%                  & -8.6        &  94         & 169          & 0.190          \\
  \cmidrule(lr){1-7}
  \multirow{6}{*}{Gemini 3 Flash Preview}
    & Model CUA                                              & 50.0\%                  & -19.9         & 40          & 33           & 0.040          \\
    & AppEvalPilot                                           & 69.9\%                  & ---         & 114         & 186          & 0.070          \\
    & UI-TARS                                                & 54.8\%                  & -15.1       &  84         & 143          & \textbf{0.020} \\
    & WebVoyager                                             & 68.9\%                  & -1.0        &  90         & 169          & 0.035          \\
  \rowcolor{blue!6}
    & \textbf{\textsc{DiagEval}} ($\times 1$)                & 76.1\%                  & +6.2        &  62         & 166          & 0.122          \\
  \rowcolor{blue!6}
    & \textbf{\textsc{DiagEval}} ($\times 2$)                & \textbf{78.3\%}         & \textbf{+8.4} & \textbf{46} & 160          & 0.154          \\
  \midrule
  \rowcolor{black!8}
  \multicolumn{7}{c}{\emph{(b) RDB (429 cases)}} \\
  \midrule
  Qwen3.5-35B-A3B
    & Model CUA                                              & 42.2\%                  & *         & 238          & 101           & 0.09            \\
  \cmidrule(lr){1-7}
  \multirow{4}{*}{Claude Opus 4.6}
    & Model CUA                                              & 69.0\%                  & +1.0      & 80          & 52            & 0.35            \\
    & AppEvalPilot                                           & 69.9\%                  & ---       & 104         & 66            & 0.385          \\
    & UI-TARS                                                & 71.3\%                  & +1.4      & 97          & 65            & 0.188          \\
    & WebVoyager                                             & 70.4\%                  & +0.5      & 83          & 53            & 0.188          \\
  \cmidrule(lr){1-7}
  \multirow{6}{*}{Gemini 3 Flash Preview}
    & Model CUA                                              & 62.2\%                  & -2.8      & 50          & 36            & 0.019          \\
    & AppEvalPilot                                           & 65.0\%                  & ---       & 132         & \textbf{73}   & 0.112          \\
    & UI-TARS                                                & 52.9\%                  & -12.1     & 181         & 71            & \textbf{0.02} \\
    & WebVoyager                                             & 68.3\%                  & +3.3      & 77          & 62            & 0.025          \\
  \rowcolor{blue!6}
    & \textbf{\textsc{DiagEval}} ($\times 1$)                & 78.6\%                  & +13.6     & 70          & 69            & 0.183          \\
  \rowcolor{blue!6}
    & \textbf{\textsc{DiagEval}} ($\times 2$)                & \textbf{81.6\%}         & \textbf{+16.6} & 50      & 62            & 0.234          \\
  \bottomrule
  \end{tabular*}
  \vspace{-1.5em}
\end{table*}

\vspace{-0.8em}
\subsection{Ablation Study}
\label{sec:repair}
\vspace{-0.5em}

\par\noindent\textbf{Diagnostic-Mechanism Ablation.}
\label{sec:repair_components}
To isolate recovery effects, we evaluate all methods on the \emph{same FN subsets}. Attribution evidence is measured by $\Delta H = H_{\mathrm{bin}}(p^{(0)}) - H_{\mathrm{bin}}(p_{\mathrm{end}})$, the entropy drop from the neutral signal $p^{(0)}=0.5$ to the terminal signal $p_{\mathrm{end}}$. This analysis focuses on FN recovery, while bilateral attribution quality is deferred later. We compare \textbf{NR}, a naive rerun; \textbf{NR+IE}, which adds FDS context and fork-point localization but removes SOU-guided branching and attribution updates; and full \textsc{DiagEval}, which further adds targeted diagnostic branching.
Table~\ref{tab:retry_ablation} shows that FN recovery and $\Delta H$ improve together only with SOU-guided branching. NR+IE provides modest gains over naive rerun with low $\Delta H$, whereas full \textsc{DiagEval} substantially improves both metrics: on WDJ-U, \textsc{DiagEval} ($\times 1$) more than doubles recovery over NR+IE (45.6\% vs. 20.2\%) and quadruples $\Delta H$ (0.274 vs. 0.063). \textsc{DiagEval} ($\times 1$) also outperforms the two-round NR and NR+IE ablations, indicating that the gain comes from diagnosis quality rather than additional budget. For budget-matched naive sampling on WDJ-U FN cases, \textsc{DiagEval} ($\times 2$) reaches 59.6\% recovery, a 106.2\% relative gain over the optimistic Best-of-3 baseline (28.9\%) in Table~\ref{tab:appendix_naive_retry_wdju}.

\par\noindent\textbf{Bilateral Attribution Score: Magnitude and Quality.}
\label{sec:bilateral}
We next test whether the attribution signal ranks failure types beyond the final pass/fail verdict. As shown in Table~\ref{tab:diageval_bilateral}, on resolved cases the terminal signal $p_{\mathrm{end}}$ aligns with the ground-truth attribution: recovered FN cases settle at low $p_{\mathrm{end}}$, attributing toward \textsc{AgentFail}, whereas retained TN cases settle at high $p_{\mathrm{end}}$, attributing toward \textsc{EnvFail}; missed cases exhibit the complementary pattern. Using $p_{\mathrm{end}}$ as the ranking score, \textsc{DiagEval} attains ROC-AUC values of $0.620/0.720$ at $\times1$ and $0.761/0.763$ at $\times2$ on RDB/WDJ-U. The signal therefore acts as a bilateral attribution score beyond the final verdict, although it is not a calibrated probability.

\vspace{-0.1em}
\begin{table*}[!t]
\centering
\caption{Component ablation on FN subsets. \textit{FDS}: failure diagnosis; \textit{Fork}: fork-point resume; \textit{Branch+Att.}: SOU-guided branching with attribution score update. \textit{Rec.\%}: FN recovery rate; \textit{Cost}: USD per case; $\Delta H$: entropy reduction in the internal attribution score from a neutral prior (higher indicates stronger attribution evidence). Best within each budget block in bold.}
\label{tab:retry_ablation}
\footnotesize
\setlength{\tabcolsep}{1.2pt}\renewcommand{\arraystretch}{0.84}
\begin{tabular*}{\textwidth}{@{\extracolsep{\fill}}lcccccccccc@{}}
\toprule
& \multicolumn{3}{c}{\textbf{Components}} & \multicolumn{3}{c}{\textbf{RDB} ($n{=}132$)} & \multicolumn{3}{c}{\textbf{WDJ-U} ($n{=}114$)} \\
\cmidrule(lr){2-4} \cmidrule(lr){5-7} \cmidrule(lr){8-10}
\textbf{Method} & FDS & Fork & Branch+Att. & Rec.\% $\uparrow$ & Cost $\downarrow$ & $\Delta H$ $\uparrow$ & Rec.\% $\uparrow$ & Cost $\downarrow$ & $\Delta H$ $\uparrow$ \\
\midrule
\textsc{DiagEval} w/o Diagnose (NR, $\times$1)     & --         & --         & --         & 32.6          & 0.030 & 0.001          & 17.5          & 0.059 & 0.046          \\
\textsc{DiagEval} w/o Branch (NR+IE, $\times$1)    & \checkmark & \checkmark & --         & 37.1          & 0.018 & 0.024          & 20.2          & 0.053 & 0.063          \\
\textsc{DiagEval} ($\times$1)                       & \checkmark & \checkmark & \checkmark & \textbf{47.0} & 0.183 & \textbf{0.305} & \textbf{45.6} & 0.123 & \textbf{0.274} \\
\midrule
\textsc{DiagEval} w/o Diagnose (NR, $\times$2)     & --         & --         & --         & 46.2          & 0.057 & 0.012          & 28.1          & 0.115 & 0.120          \\
\textsc{DiagEval} w/o Branch (NR+IE, $\times$2)    & \checkmark & \checkmark & --         & 43.2          & 0.045 & 0.024          & 30.7          & 0.130 & 0.141          \\
\textsc{DiagEval} ($\times$2)                       & \checkmark & \checkmark & \checkmark & \textbf{62.1} & 0.234 & \textbf{0.325} & \textbf{59.6} & 0.163 & \textbf{0.410} \\
\bottomrule
\end{tabular*}
\end{table*}

\begin{table}[H]
\centering
\caption{\textbf{Score-based bilateral attribution by \textsc{DiagEval}.}
Resolved cases are FN examples recovered to \textit{pass} or TN examples retained as \textit{fail}; missed cases are the corresponding complements. We report the mean score assigned to the ground-truth attribution direction and the ROC-AUC of $P(\text{env})$ for ranking TN over FN cases.}

\label{tab:diageval_bilateral}
\footnotesize
\setlength{\tabcolsep}{2.9pt}
\renewcommand{\arraystretch}{0.88}
\begin{tabular}{@{}lccccc@{}}
\toprule
& \multicolumn{2}{c}{\textbf{FN}: GT$=\textsc{AgentFail}$} & \multicolumn{2}{c}{\textbf{TN}: GT$=\textsc{EnvFail}$} & \\
\cmidrule(lr){2-3}\cmidrule(lr){4-5}
\textbf{Setting} & $\bar{P}(\text{agent})\mid_{\text{res}}$ $\uparrow$ & $\bar{P}(\text{agent})\mid_{\text{miss}}$ $\downarrow$ & $\bar{P}(\text{env})\mid_{\text{res}}$ $\uparrow$ & $\bar{P}(\text{env})\mid_{\text{miss}}$ $\downarrow$ & \textbf{ROC-AUC} $\uparrow$ \\
\midrule
\multicolumn{6}{@{}l}{\emph{RealDevBench (RDB)}} \\
\textsc{DiagEval} ($\times$1) & 0.670 & \textbf{0.126} & \textbf{0.805} & 0.430 & 0.620 \\
\textsc{DiagEval} ($\times$2) & \textbf{0.672} & 0.208 & 0.803 & \textbf{0.374} & \textbf{0.761} \\
\midrule
\multicolumn{6}{@{}l}{\emph{WebDevJudge-Unit (WDJ-U)}} \\
\textsc{DiagEval} ($\times$1) & \textbf{0.659} & \textbf{0.135} & \textbf{0.873} & \textbf{0.402} & 0.720 \\
\textsc{DiagEval} ($\times$2) & 0.657 & 0.148 & 0.864 & 0.431 & \textbf{0.763} \\
\bottomrule
\end{tabular}
\vspace{-0.6em}
\end{table}

\par\noindent\textbf{Efficiency Analysis of Branch Ordering.}
We further assess EIG-based diagnostic efficiency by comparing branch ordering methods. As detailed in Appendix~\ref{app:branch_prioritization_results}, EIG ordering achieves the highest first-branch success rate ($66\%$) and the lowest probe cost ($1.48$ executed probes per recoverable case), showing a clear efficiency advantage over prompt-based ordering methods.

\vspace{-0.8em}
\subsection{Cross-Framework Transfer}
\label{sec:transfer}
\vspace{-0.5em}
We evaluate whether the proposed diagnostic mechanism transfers across GUI-agent frameworks, with details provided in Appendix~\ref{app:transfer}. We use UI-TARS as the GUI-agent evaluation framework and invoke \textsc{DiagEval}'s online diagnosis procedure when UI-TARS returns an initial \textsc{Fail} verdict. \textsc{DiagEval} achieves a maximum absolute end-to-end accuracy improvement of $22.1$ and $21.7$ percentage points on WDJ-U and RDB, respectively. These gains indicate that \textsc{DiagEval} operates as a framework-agnostic evaluator-side mechanism, rather than a prompt-level enhancement tied to a specific GUI-agent framework.
\section{Conclusion and Limitations}
\label{sec:conclusion}
\vspace{-0.6em}

\textsc{DiagEval} shows that post-failure GUI evaluation is better framed as trajectory-conditioned diagnosis than as repeated retry. By reusing prior trajectories, launching targeted probes, and aggregating cross-trajectory evidence, it recovers 45.6--62.1\% of false negatives (vs.\ 17.5--46.2\% for retry), yields mean entropy reduction of $\Delta H = 0.27$--$0.41$ over the binary attribution score under the fixed update model, and lifts full-set accuracy from 69.9\% to 78.3\% on WebDevJudge-Unit and from 65.0\% to 81.6\% on RealDevBench. More broadly, when correctness is graph-level but evidence is path-local, reliability depends on adaptive evidence acquisition rather than attempt multiplicity alone.

\paragraph{Limitations.}
\textsc{DiagEval} relies on hand-tuned likelihood hyperparameters, prompt-implemented branch ranking, and a binary attribution space, so the resulting score is an internal diagnostic signal rather than a calibrated posterior. The fail-update rule also assumes $\Prob(\text{fail}\mid\textsc{EnvFail},d)=1$ for every probe type, which may be too strong. The diagnostic loop further depends on an external LLM supervisor $\mathcal{J}$ for FDS construction and branch generation (see \S\ref{sec:setup}), so failures or biases of $\mathcal{J}$ can propagate into both probe selection and the resulting attribution signal. Future work should learn calibrated likelihoods, reduce prompt dependence in branch selection, study cross-supervisor robustness, and extend attribution beyond the binary split.

\bibliographystyle{plainnat}
\bibliography{references}

@inproceedings{zhou2024webarena,
  title={WebArena: A Realistic Web Environment for Building Autonomous Agents},
  author={Zhou, Shuyan and Xu, Frank F and Zhu, Hao and Zhou, Xuhui and Lo, Robert and Sridhar, Abishek and Cheng, Xianyi and Ou, Tianyue and Bisk, Yonatan and Fried, Daniel and Alon, Uri and Neubig, Graham},
  booktitle={Proceedings of the International Conference on Learning Representations (ICLR)},
  year={2024}
}

@inproceedings{xie2024osworld,
  title={{OSW}orld: Benchmarking Multimodal Agents for Open-Ended Tasks in Real Computer Environments},
  author={Tianbao Xie and Danyang Zhang and Jixuan Chen and Xiaochuan Li and Siheng Zhao and Ruisheng Cao and Toh Jing Hua and Zhoujun Cheng and Dongchan Shin and Fangyu Lei and Yitao Liu and Yiheng Xu and Shuyan Zhou and Silvio Savarese and Caiming Xiong and Victor Zhong and Tao Yu},
  booktitle={Advances in Neural Information Processing Systems (NeurIPS)},
  year={2024},
  url={https://openreview.net/forum?id=tN61DTr4Ed}
}

@inproceedings{jimenez2024swebench,
  title={SWE-bench: Can Language Models Resolve Real-World GitHub Issues?},
  author={Jimenez, Carlos E and Yang, John and Wettig, Alexander and Yao, Shunyu and Pei, Kexin and Press, Ofir and Narasimhan, Karthik},
  booktitle={Proceedings of the International Conference on Learning Representations (ICLR)},
  year={2024}
}

@inproceedings{rawles2024androidworld,
  title     = {{AndroidWorld}: A Dynamic Benchmarking Environment for Autonomous Agents},
  author    = {Rawles, Chris and Clinckemaillie, Sarah and Chang, Yifan and Waltz, Jonathan and Lau, Gabrielle and Fair, Marybeth and Li, Alice and Bishop, William and Li, Wei and Campbell-Ajala, Folawiyo and Toyama, Daniel and Berry, Robert and Tyamagundlu, Divya and Lillicrap, Timothy and Riva, Oriana},
  booktitle = {Proceedings of the International Conference on Learning Representations},
  year      = {2025},
  url       = {https://proceedings.iclr.cc/paper_files/paper/2025/hash/01a83bc2f2732a58e6aa731e659e7101-Abstract-Conference.html}
}

@inproceedings{shinn2023reflexion,
  title={Reflexion: Language Agents with Verbal Reinforcement Learning},
  author={Shinn, Noah and Cassano, Federico and Gopinath, Ashwin and Narasimhan, Karthik and Yao, Shunyu},
  booktitle={Advances in Neural Information Processing Systems (NeurIPS)},
  year={2023}
}

@inproceedings{madaan2023selfrefine,
  title={Self-Refine: Iterative Refinement with Self-Feedback},
  author={Madaan, Aman and Tandon, Niket and Gupta, Prakhar and Hallinan, Skyler and Gao, Luyu and Wiegreffe, Sarah and Alon, Uri and Dziri, Nouha and Prabhumoye, Shrimai and Yang, Yiming and others},
  booktitle={Advances in Neural Information Processing Systems (NeurIPS)},
  year={2023}
}

@inproceedings{yao2023react,
  title={ReAct: Synergizing Reasoning and Acting in Language Models},
  author={Yao, Shunyu and Zhao, Jeffrey and Yu, Dian and Du, Nan and Shafran, Izhak and Narasimhan, Karthik and Cao, Yuan},
  booktitle={Proceedings of the International Conference on Learning Representations (ICLR)},
  year={2023}
}

@misc{chen2021humaneval,
  title={Evaluating Large Language Models Trained on Code},
  author={Mark Chen and Jerry Tworek and Heewoo Jun and Qiming Yuan and Henrique Ponde de Oliveira Pinto and Jared Kaplan and Harri Edwards and Yuri Burda and Nicholas Joseph and Greg Brockman and Alex Ray and Raul Puri and Gretchen Krueger and Michael Petrov and Heidy Khlaaf and Girish Sastry and Pamela Mishkin and Brooke Chan and Scott Gray and Nick Ryder and Mikhail Pavlov and Alethea Power and Lukasz Kaiser and Mohammad Bavarian and Clemens Winter and Philippe Tillet and Felipe Petroski Such and Dave Cummings and Matthias Plappert and Fotios Chantzis and Elizabeth Barnes and Ariel Herbert-Voss and William Hebgen Guss and Alex Nichol and Alex Paino and Nikolas Tezak and Jie Tang and Igor Babuschkin and Suchir Balaji and Shantanu Jain and William Saunders and Christopher Hesse and Andrew N. Carr and Jan Leike and Josh Achiam and Vedant Misra and Evan Morikawa and Alec Radford and Matthew Knight and Miles Brundage and Mira Murati and Katie Mayer and Peter Welinder and Bob McGrew and Dario Amodei and Sam McCandlish and Ilya Sutskever and Wojciech Zaremba},
  year={2021},
  eprint={2107.03374},
  archivePrefix={arXiv},
  primaryClass={cs.LG},
  url={https://arxiv.org/abs/2107.03374}
}

@misc{zhuge2024agent,
  title={Agent-as-a-Judge: Evaluate Agents with Agents},
  author={Zhuge, Mingchen and Zhao, Changsheng and Ashley, Dylan and Wang, Wenyi and Khizbullin, Dmitrii and Xiong, Yunyang and Liu, Zechun and Chang, Ernie and Krishnamoorthi, Raghuraman and Tian, Yuandong and Shi, Yangyang and Chandra, Vikas and Schmidhuber, J{\"u}rgen},
  year={2024},
  eprint={2410.10934},
  archivePrefix={arXiv},
  primaryClass={cs.AI},
  url={https://arxiv.org/abs/2410.10934}
}

@inproceedings{lu2025webgenbench,
  title={WebGen-Bench: Evaluating LLMs on Generating Interactive and Functional Websites from Scratch},
  author={Lu, Zimu and Yang, Yunqiao and Ren, Houxing and Hou, Haotian and Xiao, Han and Wang, Ke and Shi, Weikang and Zhou, Aojun and Zhan, Mingjie and Li, Hongsheng},
  booktitle={Advances in Neural Information Processing Systems (NeurIPS)},
  year={2025},
  eprint={2505.03733},
  archivePrefix={arXiv},
  primaryClass={cs.CL},
  url={https://arxiv.org/abs/2505.03733}
}

@inproceedings{li2026webdevjudge,
  title     = {{WebDevJudge}: Evaluating {(M)LLMs} as Critiques for Web Development Quality},
  author    = {Li, Chunyang and Zheng, Yilun and Huang, Xinting and Fang, Tianqing and Xu, Jiahao and Chen, Lihui and Song, Yangqiu and Hu, Han},
  booktitle = {Proceedings of the International Conference on Learning Representations},
  year      = {2026},
  url       = {https://openreview.net/forum?id=CCSPm6V5EF}
}

@inproceedings{xiao2025codeaesthetics,
  title     = {Code Aesthetics with Agentic Reward Feedback},
  author    = {Xiao, Bang and Jiang, Lingjie and Huang, Shaohan and Lv, Tengchao and Huang, Yupan and Wu, Xun and Cui, Lei and Wei, Furu},
  booktitle = {Proceedings of the International Conference on Learning Representations},
  year      = {2026},
  url       = {https://openreview.net/forum?id=Q87kwGI6bx}
}

@article{kong2026webtestbench,
  title={WebTestBench: Evaluating Computer-Use Agents towards End-to-End Automated Web Testing},
  author={Kong, Fanheng and Zhang, Jingyuan and Yue, Yang and Sun, Chenxi and Tian, Yang and Feng, Shi and Yang, Xiaocui and Wang, Daling and Tian, Yu and Du, Jun and Zeng, Wenchong and Li, Han and Gai, Kun},
  journal={arXiv preprint arXiv:2603.25226},
  year={2026}
}

@inproceedings{huang2024selfcorrect,
  title={Large Language Models Cannot Self-Correct Reasoning Yet},
  author={Huang, Jie and Chen, Xinyun and Mishra, Swaroop and Zheng, Huaixiu Steven and Yu, Adams Wei and Song, Xinying and Zhou, Denny},
  booktitle={Proceedings of the International Conference on Learning Representations (ICLR)},
  year={2024}
}

@inproceedings{xu-etal-2023-instructscore,
  title = "{INSTRUCTSCORE}: Towards Explainable Text Generation Evaluation with Automatic Feedback",
  author = "Xu, Wenda  and
    Wang, Danqing  and
    Pan, Liangming  and
    Song, Zhenqiao  and
    Freitag, Markus  and
    Wang, William  and
    Li, Lei",
  editor = "Bouamor, Houda  and
    Pino, Juan  and
    Bali, Kalika",
  booktitle = "Proceedings of the 2023 Conference on Empirical Methods in Natural Language Processing",
  month = dec,
  year = "2023",
  address = "Singapore",
  publisher = "Association for Computational Linguistics",
  url = "https://aclanthology.org/2023.emnlp-main.365/",
  doi = "10.18653/v1/2023.emnlp-main.365",
  pages = "5967--5994"
}

@misc{chen2025trainingllmbasedagentssynthetic,
  title={Training LLM-Based Agents with Synthetic Self-Reflected Trajectories and Partial Masking},
  author={Yihan Chen and Benfeng Xu and Xiaorui Wang and Yongdong Zhang and Zhendong Mao},
  year={2025},
  eprint={2505.20023},
  archivePrefix={arXiv},
  primaryClass={cs.CL},
  url={https://arxiv.org/abs/2505.20023}
}

@misc{zhang2025miragebenchllmagenthallucinating,
  title={MIRAGE-Bench: LLM Agent is Hallucinating and Where to Find Them},
  author={Weichen Zhang and Yiyou Sun and Pohao Huang and Jiayue Pu and Heyue Lin and Dawn Song},
  year={2025},
  eprint={2507.21017},
  archivePrefix={arXiv},
  primaryClass={cs.AI},
  url={https://arxiv.org/abs/2507.21017}
}

@misc{ye2025aiagentswebtesting,
  title={AI Agents for Web Testing: A Case Study in the Wild},
  author={Naimeng Ye and Xiao Yu and Ruize Xu and Tianyi Peng and Zhou Yu},
  year={2025},
  eprint={2509.05197},
  archivePrefix={arXiv},
  primaryClass={cs.SE},
  url={https://arxiv.org/abs/2509.05197}
}

@misc{lu2025webgenagentenhancinginteractivewebsite,
  title={WebGen-Agent: Enhancing Interactive Website Generation with Multi-Level Feedback and Step-Level Reinforcement Learning},
  author={Zimu Lu and Houxing Ren and Yunqiao Yang and Ke Wang and Zhuofan Zong and Junting Pan and Mingjie Zhan and Hongsheng Li},
  year={2025},
  eprint={2509.22644},
  archivePrefix={arXiv},
  primaryClass={cs.CL},
  url={https://arxiv.org/abs/2509.22644}
}

@misc{garousi2025aipoweredsoftwaretestingtools,
  title={AI-powered software testing tools: A systematic review and empirical assessment of their features and limitations},
  author={Vahid Garousi and Nithin Joy and Zafar Jafarov and Alper Bu{\u{g}}ra Kele{\c{s}} and Sevde De{\u{g}}irmenci and Ece {\"O}zdemir and Ryan Zarringhalami},
  year={2024},
  eprint={2409.00411},
  archivePrefix={arXiv},
  primaryClass={cs.SE},
  url={https://arxiv.org/abs/2409.00411}
}

@INPROCEEDINGS{11186079,
  author={Kaynak, Erg{\"u}n Batuhan and Lami, Mayasah and Moslemi, Sahand and Koyuncu, Anil},
  booktitle={2025 IEEE International Conference on Software Maintenance and Evolution (ICSME)},
  title={LLMShot: Reducing Snapshot Testing Maintenance via LLMs},
  year={2025},
  volume={},
  number={},
  pages={827-832},
  doi={10.1109/ICSME64153.2025.00087}
}

@inproceedings{yang-etal-2025-aria,
  title = "Aria-{UI}: Visual Grounding for {GUI} Instructions",
  author = "Yang, Yuhao  and
    Wang, Yue  and
    Li, Dongxu  and
    Luo, Ziyang  and
    Chen, Bei  and
    Huang, Chao  and
    Li, Junnan",
  editor = "Che, Wanxiang  and
    Nabende, Joyce  and
    Shutova, Ekaterina  and
    Pilehvar, Mohammad Taher",
  booktitle = "Findings of the Association for Computational Linguistics: ACL 2025",
  month = jul,
  year = "2025",
  address = "Vienna, Austria",
  publisher = "Association for Computational Linguistics",
  url = "https://aclanthology.org/2025.findings-acl.1152/",
  doi = "10.18653/v1/2025.findings-acl.1152",
  pages = "22418--22433",
  ISBN = "979-8-89176-256-5"
}

@inproceedings{chen-etal-2025-guicourse,
  title = "{GUIC}ourse: From General Vision Language Model to Versatile {GUI} Agent",
  author = "Chen, Wentong  and
    Cui, Junbo  and
    Hu, Jinyi  and
    Qin, Yujia  and
    Fang, Junjie  and
    Zhao, Yue  and
    Wang, Chongyi  and
    Liu, Jun  and
    Chen, Guirong  and
    Huo, Yupeng  and
    Yao, Yuan  and
    Lin, Yankai  and
    Liu, Zhiyuan  and
    Sun, Maosong",
  editor = "Che, Wanxiang  and
    Nabende, Joyce  and
    Shutova, Ekaterina  and
    Pilehvar, Mohammad Taher",
  booktitle = "Proceedings of the 63rd Annual Meeting of the Association for Computational Linguistics (Volume 1: Long Papers)",
  month = jul,
  year = "2025",
  address = "Vienna, Austria",
  publisher = "Association for Computational Linguistics",
  url = "https://aclanthology.org/2025.acl-long.1065/",
  doi = "10.18653/v1/2025.acl-long.1065",
  pages = "21936--21959",
  ISBN = "979-8-89176-251-0"
}

@misc{lee2025reguidedataefficientgui,
  title={ReGUIDE: Data Efficient GUI Grounding via Spatial Reasoning and Search},
  author={Hyunseok Lee and Jeonghoon Kim and Beomjun Kim and Jihoon Tack and Chansong Jo and Jaehong Lee and Cheonbok Park and Sookyo In and Jinwoo Shin and Kang Min Yoo},
  year={2025},
  eprint={2505.15259},
  archivePrefix={arXiv},
  primaryClass={cs.LG},
  url={https://arxiv.org/abs/2505.15259}
}

@misc{zhu2025judgelmfinetunedlargelanguage,
  title     = {{JudgeLM}: Fine-tuned Large Language Models are Scalable Judges},
  author    = {Zhu, Lianghui and Wang, Xinggang and Wang, Xinlong},
  booktitle = {Proceedings of the International Conference on Learning Representations},
  year      = {2025},
  eprint    = {2310.17631},
  archivePrefix = {arXiv},
  primaryClass  = {cs.CL},
  url       = {https://arxiv.org/abs/2310.17631}
}

@misc{chan2023chatevalbetterllmbasedevaluators,
  title={ChatEval: Towards Better LLM-based Evaluators through Multi-Agent Debate},
  author={Chi-Min Chan and Weize Chen and Yusheng Su and Jianxuan Yu and Wei Xue and Shanghang Zhang and Jie Fu and Zhiyuan Liu},
  year={2023},
  eprint={2308.07201},
  archivePrefix={arXiv},
  primaryClass={cs.CL},
  url={https://arxiv.org/abs/2308.07201}
}

@misc{wei2025rocketevalefficientautomatedllm,
  title={RocketEval: Efficient Automated LLM Evaluation via Grading Checklist},
  author={Tianjun Wei and Wei Wen and Ruizhi Qiao and Xing Sun and Jianghong Ma},
  year={2025},
  eprint={2503.05142},
  archivePrefix={arXiv},
  primaryClass={cs.CL},
  url={https://arxiv.org/abs/2503.05142}
}

@inproceedings{thakur-etal-2025-judging,
  title = "Judging the Judges: Evaluating Alignment and Vulnerabilities in {LLM}s-as-Judges",
  author = "Thakur, Aman Singh  and
    Choudhary, Kartik  and
    Ramayapally, Venkat Srinik  and
    Vaidyanathan, Sankaran  and
    Hupkes, Dieuwke",
  editor = "Arviv, Ofir  and
    Clinciu, Miruna  and
    Dhole, Kaustubh  and
    Dror, Rotem  and
    Gehrmann, Sebastian  and
    Habba, Eliya  and
    Itzhak, Itay  and
    Mille, Simon  and
    Perlitz, Yotam  and
    Santus, Enrico  and
    Sedoc, Jo{\~a}o  and
    Shmueli Scheuer, Michal  and
    Stanovsky, Gabriel  and
    Tafjord, Oyvind",
  booktitle = "Proceedings of the Fourth Workshop on Generation, Evaluation and Metrics (GEM{\texttwosuperior})",
  month = jul,
  year = "2025",
  address = "Vienna, Austria and virtual meeting",
  publisher = "Association for Computational Linguistics",
  url = "https://aclanthology.org/2025.gem-1.33/",
  pages = "404--430",
  ISBN = "979-8-89176-261-9"
}

@misc{li2025llmsreliablyjudgeyet,
  title={LLMs Cannot Reliably Judge (Yet?): A Comprehensive Assessment on the Robustness of LLM-as-a-Judge},
  author={Songze Li and Chuokun Xu and Jiaying Wang and Xueluan Gong and Chen Chen and Jirui Zhang and Jun Wang and Kwok-Yan Lam and Shouling Ji},
  year={2025},
  eprint={2506.09443},
  archivePrefix={arXiv},
  primaryClass={cs.CR},
  url={https://arxiv.org/abs/2506.09443}
}

@inproceedings{chen2024mllmasajudge,
  title={{MLLM}-as-a-Judge: Assessing Multimodal {LLM}-as-a-Judge with Vision-Language Benchmark},
  author={Dongping Chen and Ruoxi Chen and Shilin Zhang and Yaochen Wang and Yinuo Liu and Huichi Zhou and Qihui Zhang and Yao Wan and Pan Zhou and Lichao Sun},
  booktitle={Forty-first International Conference on Machine Learning},
  year={2024},
  url={https://openreview.net/forum?id=dbFEFHAD79}
}

@misc{ye2024justiceprejudicequantifyingbiases,
  title={Justice or Prejudice? Quantifying Biases in LLM-as-a-Judge},
  author={Jiayi Ye and Yanbo Wang and Yue Huang and Dongping Chen and Qihui Zhang and Nuno Moniz and Tian Gao and Werner Geyer and Chao Huang and Pin-Yu Chen and Nitesh V Chawla and Xiangliang Zhang},
  year={2024},
  eprint={2410.02736},
  archivePrefix={arXiv},
  primaryClass={cs.CL},
  url={https://arxiv.org/abs/2410.02736}
}

@inproceedings{NEURIPS2023_91f18a12,
  title = {Judging {{LLM-as-a-judge}} with {{MT-bench}} and Chatbot Arena},
  booktitle = {Advances in Neural Information Processing Systems},
  author = {Zheng, Lianmin and Chiang, Wei-Lin and Sheng, Ying and Zhuang, Siyuan and Wu, Zhanghao and Zhuang, Yonghao and Lin, Zi and Li, Zhuohan and Li, Dacheng and Xing, Eric and Zhang, Hao and Gonzalez, Joseph E and Stoica, Ion},
  editor = {Oh, A. and Naumann, T. and Globerson, A. and Saenko, K. and Hardt, M. and Levine, S.},
  year = {2023},
  volume = {36},
  pages = {46595--46623},
  publisher = {Curran Associates, Inc.},
  url = {https://proceedings.neurips.cc/paper_files/paper/2023/file/91f18a1287b398d378ef22505bf41832-Paper-Datasets_and_Benchmarks.pdf}
}

@article{chan2024mle,
  title={Mle-bench: Evaluating machine learning agents on machine learning engineering},
  author={Chan, Jun Shern and Chowdhury, Neil and Jaffe, Oliver and Aung, James and Sherburn, Dane and Mays, Evan and Starace, Giulio and Liu, Kevin and Maksin, Leon and Patwardhan, Tejal and others},
  journal={arXiv preprint arXiv:2410.07095},
  year={2024}
}

@article{cheng2024seeclick,
  author  = {Cheng, Kanzhi and Sun, Qiushi and Chu, Yougang and Xu, Fangzhi and Li, Yantao and Zhang, Jianbing and Wu, Zhiyong},
  journal = {arXiv preprint arXiv:2401.10935},
  title   = {Seeclick: Harnessing gui grounding for advanced visual gui agents},
  year    = {2024}
}

@article{ding2023crosscodeeval,
  author  = {Ding, Yangruibo and Wang, Zijian and Ahmad, Wasi and Ding, Hantian and Tan, Ming and Jain, Nihal and Ramanathan, Murali Krishna and Nallapati, Ramesh and Bhatia, Parminder and Roth, Dan and others},
  journal = {Advances in Neural Information Processing Systems},
  pages   = {46701--46723},
  title   = {Crosscodeeval: A diverse and multilingual benchmark for cross-file code completion},
  volume  = {36},
  year    = {2023}
}

@inproceedings{jain2025livecodebench,
  author    = {Naman Jain and King Han and Alex Gu and Wen-Ding Li and Fanjia Yan and Tianjun Zhang and Sida Wang and Armando Solar-Lezama and Koushik Sen and Ion Stoica},
  booktitle = {The Thirteenth International Conference on Learning Representations},
  title     = {LiveCodeBench: Holistic and Contamination Free Evaluation of Large Language Models for Code},
  url       = {https://openreview.net/forum?id=chfJJYC3iL},
  year      = {2025}
}

@misc{anthropic2025harnesses,
  title        = {Effective harnesses for long-running agents},
  author       = {{Anthropic Engineering Team}},
  howpublished = {\url{https://www.anthropic.com/engineering/effective-harnesses-for-long-running-agents}},
  year         = {2025},
  note         = {Accessed: 2025-12}
}

@inproceedings{liu2024repobench,
  author    = {Tianyang Liu and Canwen Xu and Julian McAuley},
  booktitle = {The Twelfth International Conference on Learning Representations},
  title     = {RepoBench: Benchmarking Repository-Level Code Auto-Completion Systems},
  url       = {https://openreview.net/forum?id=pPjZIOuQuF},
  year      = {2024}
}

@article{miserendino2025swe,
  author  = {Miserendino, Samuel and Wang, Michele and Patwardhan, Tejal and Heidecke, Johannes},
  journal = {arXiv preprint arXiv:2502.12115},
  title   = {SWE-Lancer: Can Frontier LLMs Earn \$1 Million from Real-World Freelance Software Engineering?},
  year    = {2025}
}

@article{qin2025uitars,
  author  = {Qin, Yujia and Ye, Yining and Fang, Junjie and Wang, Haoming and Liang, Shihao and Tian, Shizuo and Zhang, Junda and Li, Jiahao and Li, Yunxin and Huang, Shijue and others},
  journal = {arXiv preprint arXiv:2501.12326},
  title   = {UI-TARS: Pioneering Automated GUI Interaction with Native Agents},
  year    = {2025}
}

@inproceedings{zhang2024naturalcodebench,
  author    = {Zhang, Shudan and Zhao, Hanlin and Liu, Xiao and Zheng, Qinkai and Qi, Zehan and Gu, Xiaotao and Dong, Yuxiao and Tang, Jie},
  booktitle = {Findings of the Association for Computational Linguistics ACL 2024},
  pages     = {7907--7928},
  title     = {Naturalcodebench: Examining coding performance mismatch on humaneval and natural user queries},
  year      = {2024}
}

@article{zhuo2024bigcodebench,
  author  = {Zhuo, Terry Yue and Vu, Minh Chien and Chim, Jenny and Hu, Han and Yu, Wenhao and Widyasari, Ratnadira and Yusuf, Imam Nur Bani and Zhan, Haolan and He, Junda and Paul, Indraneil and others},
  journal = {arXiv preprint arXiv:2406.15877},
  title   = {Bigcodebench: Benchmarking code generation with diverse function calls and complex instructions},
  year    = {2024}
}

@misc{anthropic2024claude35sonnet,
  author       = {Anthropic},
  title        = {Claude 3.5 Sonnet},
  howpublished = {\url{https://www.anthropic.com/news/claude-3-5-sonnet}},
  year         = {2024},
  note         = {Accessed on March 28, 2025.}
}

@article{he2024webvoyager,
  title={Webvoyager: Building an end-to-end web agent with large multimodal models},
  author={He, Hongliang and Yao, Wenlin and Ma, Kaixin and Yu, Wenhao and Dai, Yong and Zhang, Hongming and Lan, Zhenzhong and Yu, Dong},
  journal={arXiv preprint arXiv:2401.13919},
  year={2024}
}

@article{wu2024atlas,
  title={Os-atlas: A foundation action model for generalist gui agents},
  author={Wu, Zhiyong and Wu, Zhenyu and Xu, Fangzhi and Wang, Yian and Sun, Qiushi and Jia, Chengyou and Cheng, Kanzhi and Ding, Zichen and Chen, Liheng and Liang, Paul Pu and others},
  journal={arXiv preprint arXiv:2410.23218},
  year={2024}
}

@article{xu2024aguvis,
  title={Aguvis: Unified pure vision agents for autonomous gui interaction},
  author={Xu, Yiheng and Wang, Zekun and Wang, Junli and Lu, Dunjie and Xie, Tianbao and Saha, Amrita and Sahoo, Doyen and Yu, Tao and Xiong, Caiming},
  journal={arXiv preprint arXiv:2412.04454},
  year={2024}
}

@article{gou2024navigating,
  title={Navigating the Digital World as Humans Do: Universal Visual Grounding for GUI Agents},
  author={Gou, Boyu and Wang, Ruohan and Zheng, Boyuan and Xie, Yanan and Chang, Cheng and Shu, Yiheng and Sun, Huan and Su, Yu},
  journal={arXiv preprint arXiv:2410.05243},
  year={2024},
  eprint={2410.05243},
  archivePrefix={arXiv},
  primaryClass={cs.AI},
  url={https://arxiv.org/abs/2410.05243}
}

@inproceedings{lu2025uxagent,
  title={UXAgent: An LLM Agent-Based Usability Testing Framework for Web Design},
  author={Lu, Yuxuan and Yao, Bingsheng and Gu, Hansu and Huang, Jing and Wang, Zheshen Jessie and Li, Yang and Gesi, Jiri and He, Qi and Li, Toby Jia-Jun and Wang, Dakuo},
  booktitle={Proceedings of the Extended Abstracts of the CHI Conference on Human Factors in Computing Systems},
  pages={1--12},
  year={2025},
  publisher={Association for Computing Machinery},
  doi={10.1145/3706599.3719729},
  url={https://doi.org/10.1145/3706599.3719729},
  series={CHI EA '25}
}

@misc{google2025gemini25cum,
  title     = {Gemini 2.5 Computer Use Model},
  author    = {Google DeepMind},
  url       = {https://blog.google/technology/google-deepmind/gemini-computer-use-model},
  year      = {2025}
}

@misc{PlayCoder2026,
  title         = {{PlayCoder}: Making LLM-Generated GUI Code Playable},
  author        = {Peng, Zhiyuan and Tao, Wei and Yin, Xin and Ying, Chenhao and Luo, Yuan and Guo, Yiwen},
  year          = {2026},
  eprint        = {2604.19742},
  archivePrefix = {arXiv},
  primaryClass  = {cs.SE},
  url           = {https://arxiv.org/abs/2604.19742}
}

@article{rosset2026art,
  title         = {The Art of Building Verifiers for Computer Use Agents},
  author        = {Rosset, Corby and Sharma, Pratyusha and Zhao, Andrew and Gonzalez-Fernandez, Miguel and Awadallah, Ahmed},
  journal       = {arXiv preprint arXiv:2604.06240},
  year          = {2026},
  eprint        = {2604.06240},
  archivePrefix = {arXiv},
  primaryClass  = {cs.CR},
  doi           = {10.48550/arXiv.2604.06240},
  url           = {https://arxiv.org/abs/2604.06240}
}

@article{jin2026halluclear,
  title={HalluClear: Diagnosing, Evaluating and Mitigating Hallucinations in GUI Agents},
  author={Jin, Chao and Yang, Wenkui and Sun, Hao and Liao, Yuqi and Jiang, Qianyi and Zhou, Kai and Cao, Jie and He, Ran and Huang, Huaibo},
  journal={arXiv preprint arXiv:2604.17284},
  year={2026}
}

@misc{gao2026guitesterenablingguiagents,
  title={GUITester: Enabling GUI Agents for Exploratory Defect Discovery},
  author={Yifei Gao and Jiang Wu and Xiaoyi Chen and Yifan Yang and Zhe Cui and Tianyi Ma and Jiaming Zhang and Jitao Sang},
  year={2026},
  eprint={2601.04500},
  archivePrefix={arXiv},
  primaryClass={cs.AI},
  url={https://arxiv.org/abs/2601.04500}
}

@article{liu2025agent,
  title={Agent-Environment Alignment via Automated Interface Generation},
  author={Liu, Kaiming and Lei, Xuanyu and Wang, Ziyue and Li, Peng and Liu, Yang},
  journal={arXiv preprint arXiv:2505.21055},
  year={2025}
}

@misc{bian2025dontknowclickautomatedgui,
  title={You Don't Know Until You Click: Automated GUI Testing for Production-Ready Software Evaluation},
  author={Yutong Bian and Xianhao Lin and Yupeng Xie and Tianyang Liu and Mingchen Zhuge and Siyuan Lu and Haoming Tang and Jinlin Wang and Jiayi Zhang and Jiaqi Chen and Xiangru Tang and Yongxin Ni and Sirui Hong and Chenglin Wu},
  year={2025},
  eprint={2508.14104},
  archivePrefix={arXiv},
  primaryClass={cs.SE},
  url={https://arxiv.org/abs/2508.14104}
}

@article{chen2026seeing,
  title={Seeing the Whole Elephant: A Benchmark for Failure Attribution in LLM-based Multi-Agent Systems},
  author={Chen, Mengzhuo and Wang, Junjie and Mu, Fangwen and Wang, Yawen and Liu, Zhe and Feng, Huanxiang and Wang, Qing},
  journal={arXiv preprint arXiv:2604.22708},
  year={2026}
}

@misc{smirnova2019distributionallyrobustreinforcementlearning,
  title={Distributionally Robust Reinforcement Learning},
  author={Elena Smirnova and Elvis Dohmatob and J{\'e}r{\'e}mie Mary},
  year={2019},
  eprint={1902.08708},
  archivePrefix={arXiv},
  primaryClass={stat.ML},
  url={https://arxiv.org/abs/1902.08708}
}

@misc{zhang2025agentracer,
  title={AgenTracer: Who Is Inducing Failure in the LLM Agentic Systems?},
  author={Guibin Zhang and Junhao Wang and Junjie Chen and Wangchunshu Zhou and Kun Wang and Shuicheng Yan},
  year={2025},
  eprint={2509.03312},
  archivePrefix={arXiv},
  primaryClass={cs.CL},
  url={https://arxiv.org/abs/2509.03312}
}

@article{veiga2023reactive,
  title={From reactive to active sensing: A survey on information gathering in decision-theoretic planning},
  author={Veiga, Tiago and Renoux, Jennifer},
  journal={ACM Computing Surveys},
  volume={55},
  number={13s},
  pages={1--22},
  year={2023},
  publisher={ACM New York, NY}
}

@misc{qwen2026qwen35,
  title  = {{Qwen3.5}: Towards Native Multimodal Agents},
  author = {{Qwen Team}},
  month  = {February},
  year   = {2026},
  url    = {https://qwen.ai/blog?id=qwen3.5}
}

@misc{anthropic2026claudeopus46,
  title        = {{Claude Opus 4.6}},
  author       = {{Anthropic}},
  year         = {2026},
  month        = feb,
  url          = {https://www.anthropic.com/news/claude-opus-4-6},
  note         = {Anthropic model announcement and system-card reference}
}

@misc{google2025gemini3flash,
  title        = {{Gemini 3 Flash} Model Card},
  author       = {{Google DeepMind}},
  year         = {2025},
  month        = dec,
  url          = {https://storage.googleapis.com/deepmind-media/Model-Cards/Gemini-3-Flash-Model-Card.pdf},
  note         = {Model card for Gemini 3 Flash; evaluated via Gemini 3 Flash Preview API}
}

@misc{anthropic2026claudesonnet46,
  title        = {{Claude Sonnet 4.6} System Card},
  author       = {{Anthropic}},
  year         = {2026},
  month        = feb,
  url          = {https://www.anthropic.com/claude-sonnet-4-6-system-card},
  note         = {System card for Claude Sonnet 4.6}
}

\newpage
\appendix
\section{Broader Impact}

\textsc{DiagEval} improves the reliability of automated software evaluation by
separating evaluator-side errors from genuine defects. However, because GUI
agents process screenshots, DOM states, and interactive workflows, this approach
risks exposing sensitive user data. Furthermore, enhanced UI exploration
capabilities could be maliciously misused to probe software vulnerabilities. To
mitigate these risks, deployments must strictly use sandboxed environments,
exclude sensitive data, ensure fully auditable diagnostic traces, and restrict
application to authorized testing settings.
\section{Detailed SOU and Diagnostic Operators}
\label{sec:appendix_sou}

\begin{table*}[h]
\centering
\caption{Detailed sources of uncertainty (SOU) and diagnostic operators. This table expands the corruption mode and observable cue for each SOU.}
\label{tab:sou_full}
\footnotesize
\setlength{\tabcolsep}{4pt}
\renewcommand{\arraystretch}{0.98}
\begin{tabular*}{\textwidth}{@{\extracolsep{\fill}}p{2.6cm}p{4.1cm}p{5.6cm}c@{}}
\toprule
\textbf{SOU} & \textbf{Corruption} & \textbf{Cue} & \textbf{Op.} \\
\midrule
Imperfect grounding & Wrong target or actuation & A blocked edge may still be reachable & A \\
Incomplete observation & Hidden viewport, tab, or state & A continuation may lie off-screen or off-state & B \\
Reasoning hallucination & Misread UI semantics & A reported blockage may be spurious & A/C \\
Runtime instability & Non-deterministic execution & A failure may be transient & C \\
\bottomrule
\end{tabular*}
\end{table*}

\section{Sequential Diagnostic Algorithm}
\label{sec:appendix_algorithm}

\paragraph{Overall diagnostic procedure.}
The SOU operators in Table~\ref{tab:sou_full} define the evidence channels used by \textsc{DiagEval}. Algorithm~\ref{alg:diageval} summarizes how these operators are invoked in the sequential diagnostic loop. After an initial failed evaluation, \textsc{DiagEval} constructs an FDS, generates and selects diagnostic branches, orders them by EIG score, and updates the attribution score after each executed probe. Diagnosis terminates when a branch verifies success, when the attribution signal exceeds $\tau_{\mathrm{env}}$, or when the diagnostic budget is exhausted.

\vspace{0.35em}
\begin{center}
\begin{minipage}{0.99\textwidth}
\refstepcounter{customalg}
\label{alg:diageval}

\hrule height 0.8pt
\vspace{0.35em}
\noindent\textbf{Algorithm~\thecustomalg: \textsc{DiagEval} Sequential Diagnostic Evaluation}
\vspace{0.35em}
\hrule height 0.45pt
\vspace{0.45em}

\small
\noindent\textbf{Input:}
\texttt{task}; evaluator $\mathcal{E}$; rounds $R$; candidates
$N_{\mathrm{cand}}$; budget $K$; threshold $\tau_{\mathrm{env}}$.

\noindent\textbf{Output:}
final verdict $\hat{y}$ and terminal attribution score $p$.

\vspace{0.35em}
\setlength{\tabcolsep}{2pt}
\renewcommand{\arraystretch}{1.02}
\begin{tabular}{@{}r@{\hspace{0.75em}}p{0.90\textwidth}@{}}
1 &
$(\traj_1,\hat{y}_1)\leftarrow \mathcal{E}(\texttt{task})$;
if $\hat{y}_1=\textsc{Pass}$, return $\hat{y}=\textsc{Pass}$ \\

2 &
Initialize $p_{\mathrm{cur}}\leftarrow 0.5$ and $\mathcal{H}\leftarrow\emptyset$ \\

3 &
\textbf{for} $k=0,\ldots,R-1$ \textbf{do} \\

4 &
\quad $p^{(k,0)}\leftarrow p_{\mathrm{cur}}$ \\

5 &
\quad $\mathcal{F}_k=(t_k^*,h_k,D_{t_k^*},\mathcal{C}_{t_k^*})
\leftarrow \Phi_{\mathrm{FDS}}(\traj_1,\mathcal{H})$ \\

6 &
\quad $\mathcal{P}_k\leftarrow \mathcal{G}(\mathcal{F}_k,N_{\mathrm{cand}})$ \\

7 &
\quad $\widetilde{\mathcal{P}}_k
\leftarrow
\operatorname{TopK}_{K,\,b\in\mathcal{P}_k}
\mathcal{J}_{\mathrm{sel}}(b;\mathcal{F}_k,\mathcal{H})$ \\

8 &
\quad $\pi_k
\leftarrow
\operatorname{argsort}_{b\in\widetilde{\mathcal{P}}_k}
[-\mathrm{EIG}(b;\mathcal{F}_k,p^{(k,0)})]$ \\

9 &
\quad \textbf{for} $j=1,\ldots,|\pi_k|$ \textbf{do} \\

10 &
\quad\quad $b_{k,j}\leftarrow \pi_k[j]$,\quad
$d_{k,j}\leftarrow d(b_{k,j})$ \\

11 &
\quad\quad $(\traj_{k,j}^{\mathrm{br}},o_{k,j})
\leftarrow
\operatorname{Exec}(b_{k,j},\mathcal{C}_{t_k^*})$ \\

12 &
\quad\quad $p^{(k,j)}
\leftarrow
\operatorname{Update}(p^{(k,j-1)},o_{k,j},b_{k,j})$;\quad
$p_{\mathrm{cur}}\leftarrow p^{(k,j)}$ \\

13 &
\quad\quad $\mathcal{H}\leftarrow
\mathcal{H}\cup\{(b_{k,j},d_{k,j},o_{k,j},\traj_{k,j}^{\mathrm{br}},p^{(k,j)})\}$ \\

14 &
\quad\quad \textbf{if} $o_{k,j}=\texttt{verified\_success}$
\textbf{then return} $\hat{y}=\textsc{Pass}$ \\

15 &
\quad\quad \textbf{if} $p^{(k,j)}\geq\tau_{\mathrm{env}}$
\textbf{then return} $\hat{y}=\textsc{Fail}$ \\

16 &
\quad \textbf{end for} \\

17 &
\textbf{end for} \\

18 &
\textbf{return} $\hat{y}=\textsc{Fail},\ p_{\mathrm{cur}}$ \\
\end{tabular}

\vspace{0.35em}
\hrule height 0.8pt
\end{minipage}
\end{center}
\vspace{0.35em}

\section{Datasets and Metrics}
\label{sec:appendix_dataset_scope}
We evaluate on two open-source web-development benchmarks: RealDevBench (RDB)~\citep{bian2025dontknowclickautomatedgui} and WebDevJudge-Unit (WDJ-U)~\citep{li2026webdevjudge}.
RDB is a repository-level benchmark. For this dataset, we use the commercial software-generation system Atoms\footnote{\url{https://atoms.dev/}} to generate code because it provides direct run-and-deploy functionality, allowing us to collect deployed software environments without preparing offline containers.
The raw RDB annotation pool contains 662 generated cases with valid paired annotator scores after excluding construction-failure records.
From this pool, we remove cases whose deployed environments fail completely and form two disjoint subsets: a 41-case pilot set used only for lightweight sanity checks of empirical parameters, and a 429-case main test set used for all reported RDB results.
The 41-case pilot set also provides the dimension-prior counts later summarized in Tables~\ref{tab:dim_priors} and~\ref{tab:dim_prior_validation}; it is not used in the reported RDB test results.
WDJ-U contains 502 unit-level tasks with ground-truth pass/fail labels.

For both benchmarks, accuracy is the fraction of cases whose final evaluator verdict matches the ground-truth label, and average cost per case is the total API cost divided by the number of evaluated cases.
The FN subset contains cases with ground-truth \textsc{Pass} but initial verdict \textsc{Fail}; FN recovery rate is the fraction of this subset corrected to final \textsc{Pass} under the matched diagnostic setting.
For diagnostic analyses, TN flip rate measures the fraction of initially correct \textsc{Fail} verdicts on ground-truth \textsc{Fail} cases that are changed to \textsc{Pass}, and entropy reduction $\Delta H$ is computed from the internal attribution signal before and after diagnostic probing.

\section{RealDevBench Annotation Protocol}
\label{sec:appendix_annotation}

For the RealDevBench portion, we manually annotate each generated software project against the task-specific test requirements to determine whether the environment-level execution is genuinely successful.
Each case is independently reviewed by two test engineers, after which an algorithm reviewer performs calibration.
Cases with inconsistent judgments are re-executed and adjudicated in a second round.
This process yields the environment-level pass/fail labels for the RDB pilot and main-test subsets, with only the 429-case main test set used in the reported RDB results.

\paragraph{Inter-rater agreement.}
The first annotator (A) and the second annotator (B) independently assign a binary score to each case: 0 for failure and 1 for pass.
After excluding records with construction failures, 662 cases receive valid scores from both annotators.
The two annotators agree on 611 cases, yielding an observed agreement of $P_o=0.923$.
Cohen's $\kappa$ is 0.734, with an expected chance agreement of $P_e=0.710$, indicating moderate-to-substantial agreement under the standard interpretation of $\kappa$ values.
The marginal score distributions are also similar: annotator A assigns 0 to 17.37\% of cases and 1 to 82.63\%, while annotator B assigns 0 to 17.82\% and 1 to 82.18\%.
Table~\ref{tab:annotation_agreement} summarizes the confusion matrix.

\begin{table}[h]
\centering
\caption{Inter-rater agreement between the first annotator (A) and second annotator (B) after filtering construction failures. Scores are binary: 0 denotes failure and 1 denotes pass.}
\label{tab:annotation_agreement}
\small
\setlength{\tabcolsep}{10pt}
\begin{tabular}{lcc}
\toprule
 & \textbf{B = 0} & \textbf{B = 1} \\
\midrule
\textbf{A = 0} & 91 & 24 \\
\textbf{A = 1} & 27 & 520 \\
\bottomrule
\end{tabular}
\end{table}

The total disagreement rate is 7.7\% (51/662). Among the disagreements, A is stricter in 24 cases (A=0, B=1), while B is stricter in 27 cases (A=1, B=0), suggesting that the two annotators have broadly balanced strictness.

\section{Prompt Templates}
\label{sec:appendix_prompts}

This section provides the core prompt templates used by \textsc{DiagEval}'s supervisor module. All prompts produce structured JSON output for downstream processing. Full verbatim prompts are available in the supplementary code.

\subsection{Failure Diagnostic Summary (FDS) Prompt}
\label{sec:appendix_fds_prompt}

The FDS prompt takes a compressed failed trajectory and produces the initial diagnostic record.

\begin{promptbox}[coltext=PromptText,fontupper=\ttfamily\scriptsize\raggedright]{Failure Diagnostic Summary Prompt}
You are a test replay supervisor. A GUI agent just finished its first-round test run and the result is FAILURE. You are given the compressed trajectory.\par
\par
Analyze the trajectory and answer three questions in order:\par
1. **Why did it fail?** - Is it the agent's fault (wrong strategy, wrong clicks) or the environment's fault (app broken, page not loading, UI non-functional)?\par
2. **Should we retry?** - Will a second attempt with a different strategy likely succeed, or is retrying pointless?\par
3. **If retry, from which step?** - Which step is the last known-good state to restart from, and what should the agent do differently?\par
\par
\#\# How to read the trajectory\par
- **"action"** = the raw pyautogui command executed (click, scroll, key press, etc.).\par
- **"summary"** = what the agent INTENDED to do. This is NOT confirmation the action succeeded.\par
- You MUST use the **final result evidence** to retroactively judge earlier steps. If evidence says "all elements were unresponsive", then earlier clicks on those elements also failed - even if summaries sound optimistic.\par
- A click + optimistic summary != success. Only mark a step as successful if there is corroborating evidence.\par
\par
\#\# Failure type definitions\par
- **agent**: The application works (at least partially), but the agent chose wrong actions, clicked wrong elements, used a bad strategy, or gave up too early. Retry with a better approach is likely to help.\par
- **env**: The application itself is broken - page didn't load, JS not running, all UI elements non-responsive, blank/black screen, or critical infrastructure failure. Retrying the same test is unlikely to help without fixing the environment.\par
- **ambiguous**: Cannot determine clearly; some elements worked but the failure pattern doesn't clearly point to agent or environment.\par
\par
\#\# Task under test\par
\{task\_desc\}\par
\par
\#\# Trajectory\par
last\_completed\_iter: \{last\_completed\_iter\}\par
\par
\#\#\# Step-by-step timeline (action + summary pairs):\par
\{timeline\_text\}\par
\par
\#\#\# Final result:\par
\{result\_history\_text\}\par
\par
\#\# Required output\par
Return ONLY a single JSON object (no markdown fences):\par
\par
\{\par
\hspace*{1em}"failure\_type": "<agent | env | ambiguous>",\par
\hspace*{1em}"fail\_reason": "<2-3 sentences: why did the test fail? What went wrong?>",\par
\hspace*{1em}"should\_retry": <true | false>,\par
\hspace*{1em}"retry\_reason": "<1-2 sentences: why retry will/won't help>",\par
\hspace*{1em}"restart\_from\_iter": <int, 0-based step index to restart from; set to 0 if should\_retry is false or if no step succeeded>,\par
\hspace*{1em}"restart\_explanation": "<2-3 sentences: why restart from this step, and what should the agent do differently>"\par
\}\par
\par
\#\# Examples\par
\par
Agent failure (retry worthwhile):\par
\{\par
\hspace*{1em}"failure\_type": "agent",\par
\hspace*{1em}"fail\_reason": "The agent successfully loaded the page and opened the date picker, but then repeatedly clicked the same start-date field without ever selecting an end date. The date range was never completed, so the display info never updated.",\par
\hspace*{1em}"should\_retry": true,\par
\hspace*{1em}"retry\_reason": "The application is functional. The agent just needs a better strategy for selecting both start and end dates.",\par
\hspace*{1em}"restart\_from\_iter": 4,\par
\hspace*{1em}"restart\_explanation": "Step 4 successfully opened the start-date picker (confirmed by the date value changing in step 5). Restart here and immediately select a start date, then switch to the end-date field - do not re-click the start-date field."\par
\}\par
\par
Environment failure (retry not worthwhile):\par
\{\par
\hspace*{1em}"failure\_type": "env",\par
\hspace*{1em}"fail\_reason": "No UI element responded to any interaction throughout the entire session. Buttons, dropdowns, and scroll all failed. The page appears to have loaded as a static render without JavaScript execution.",\par
\hspace*{1em}"should\_retry": false,\par
\hspace*{1em}"retry\_reason": "The application is completely non-functional. Retrying with the same environment will produce the same result.",\par
\hspace*{1em}"restart\_from\_iter": 0,\par
\hspace*{1em}"restart\_explanation": "No step produced any verifiable progress. If the environment is fixed, start from scratch."\par
\}\par
\par
Ambiguous failure:\par
\{\par
\hspace*{1em}"failure\_type": "ambiguous",\par
\hspace*{1em}"fail\_reason": "The agent managed to interact with some elements (dropdown worked, scroll worked) but the core Upload File button was consistently unresponsive across 6 attempts at different coordinates.",\par
\hspace*{1em}"should\_retry": true,\par
\hspace*{1em}"retry\_reason": "Some UI elements work, so the app may be partially functional. A different upload approach (drag-and-drop, keyboard shortcut) might succeed.",\par
\hspace*{1em}"restart\_from\_iter": 2,\par
\hspace*{1em}"restart\_explanation": "Step 2 confirmed the page is partially interactive (dropdown responded). Restart here and try file upload via keyboard shortcut (Ctrl+O) or drag-and-drop instead of clicking the button."\par
\}\par
\par
\#\# Your answer (JSON only):\par
\end{promptbox}
\paragraph{Operational representation of the uncertainty profile.}
The SOU taxonomy in the main text defines conceptual evidence channels; we do not assume that each failure belongs strictly to only one SOU. In implementation, the uncertainty profile $h_k$ is represented as an operational routing profile rather than as a four-dimensional probability vector over SOUs. Concretely, we write $h_k = (c_k, w_k)$,
where $c_k$ is a discrete failure category inferred from the FDS failure reason, and $w_k$ is a normalized distribution over the three diagnostic branch dimensions $\mathcal{D}=\{A,B,C\}$. The FDS first produces structured fields such as \texttt{failure\_type}, \texttt{fail\_reason}, \texttt{restart\_from\_iter}, and \texttt{restart\_explanation}.

\subsection{Failure Category Classification Prompt}
\label{sec:appendix_category_prompt}

After the FDS is produced, a second inference pass expands the SOU-driven probing strategy into three probe types and five concrete subcategories, each corresponding to a specific diagnostic objective and intervention pattern (as mentioned in \S\ref{sec:probe_gen}).
Each category is anchored by 2--3 concrete examples in the prompt (e.g., for \texttt{wrong\_target}: ``The agent clicked the preset text label instead of the actual radio button circle, so the resize was never applied''). The classified category maps to empirical per-dimension success priors (Table~\ref{tab:dim_priors}) for branch budget allocation.

\begin{promptbox}[coltext=PromptText,fontupper=\ttfamily\scriptsize\raggedright]{Failure Category Classification Prompt}
Classify the agent's failure reason into exactly ONE category.\par
\par
\#\# Categories\par
\par
- **insufficient\_exploration**: The agent gave up too early or failed to navigate/scroll\par
enough. It drew premature conclusions without fully exploring the page or available UI.\par
\par
- **wrong\_strategy**: The agent located the correct elements but used the wrong technique,\par
algorithm, or execution approach (wrong key timing, wrong game strategy, wrong sequence).\par
\par
- **wrong\_target**: The agent interacted with the wrong element, wrong coordinates, or\par
misidentified what a UI component does. The element itself may be correct but the agent\par
pointed at the wrong thing.\par
\par
- **env\_boundary**: The feature may not be implemented, the app has a broken component,\par
or the UI is non-responsive regardless of what the agent tries.\par
\par
- **unknown**: The failure does not clearly fit any of the above categories.\par
\par
\#\# Examples\par
\par
[insufficient\_exploration]\par
"The agent only pressed 'pagedown' twice and then concluded no carousel exists. It never took a screenshot to confirm the page state."\par
"The agent completed the survey but could not find dimension percentage scores on the results page. It did not scroll through the full results page."\par
"The agent scrolled through the timeline but did not explore all UI controls or buttons that might reveal comparison photos."\par
\par
[wrong\_strategy]\par
"The agent moved the paddle briefly (0.5 seconds) and then the ball was lost. It never implemented continuous paddle control to keep the ball alive."\par
"The agent repeatedly anchored on card (0,0) paired with every other card - an invalid memory match strategy since both cards need to be different."\par
"The agent performed hard drops without attempting to clear lines, so the LINES counter remained at 0."\par
\par
[wrong\_target]\par
"The agent repeatedly attempted to fill form fields using incorrect coordinates, resulting in validation errors."\par
"The agent clicked the preset text label instead of the actual radio button circle, so the resize was never applied."\par
"The agent attempted to drag a locked (correctly-placed) puzzle piece instead of testing an unlocked piece."\par
\par
[env\_boundary]\par
"Clicking a tag on a note card navigated to the detail page instead of filtering - the tag filtering feature may not be implemented."\par
"The agent waited 105 seconds but no inactivity prompt appeared - the timeout threshold may be longer than tested or not implemented."\par
"Every time a shape property value was committed, the selected shape disappeared from the canvas - likely an application bug."\par
\par
\#\# Failure reason to classify\par
\{fail\_reason\}\par
\par
Return JSON only, no markdown:\par
\{"category": "insufficient\_exploration | wrong\_strategy | wrong\_target | env\_boundary | unknown"\}\par
\end{promptbox}

\begin{table}[h]
\centering
\caption{Empirical per-dimension success counts by failure category, used for SOU-driven dimension allocation. Source: 103 branches across the 41-case RDB pilot set.}
\label{tab:dim_priors}
\small
\begin{tabular}{lccc}
\toprule
\textbf{Failure category} & \textbf{Dim A} & \textbf{Dim B} & \textbf{Dim C} \\
\midrule
\texttt{insufficient\_exploration} & 1 & 7 & 0 \\
\texttt{wrong\_strategy} & 3 & 0 & 0 \\
\texttt{wrong\_target} & 2 & 0 & 0 \\
\texttt{env\_boundary} & 2 & 1 & 0 \\
\texttt{unknown} & 0 & 0 & 0 \\
\bottomrule
\end{tabular}
\end{table}

\paragraph{Source of dimension priors.}
The counts in Table~\ref{tab:dim_priors} are derived from the 41-case RDB pilot
set, consisting of 41 agent-fail cases and 103 executed diagnostic branches.
The table reports only successful-branch counts; failed executed branches are part
of the 103-branch source pool but do not contribute to the cells.
We use these counts as lightweight routing priors for SOU-guided
branch-dimension allocation. The 429-case RDB main test set and WDJ-U are held out
from this pilot source and have zero case-level overlap with it.

To check whether these pilot-derived priors reflect stable diagnostic structure,
we further compare the dominant successful dimension on held-out evaluation logs
in Table~\ref{tab:dim_prior_validation}. For the categories with clear pilot
signal, the dominant successful dimension is consistent across the pilot, the
RDB main-test FN set, and the WebDevJudge-Unit FN set:
\texttt{wrong\_strategy}, \texttt{wrong\_target}, and \texttt{env\_boundary}
are all dominated by Dimension~A. Thus, the pilot counts serve as coarse
dimension-level routing priors rather than case-specific statistics. We apply
Laplace smoothing before converting the counts into allocation weights, so every
diagnostic dimension retains non-zero probability.

\begin{table}[h]
\centering
\caption{\textbf{Held-out consistency of pilot-derived dimension priors.}
Each entry reports the dominant successful dimension followed by the raw
successful-branch counts along Dimensions A/B/C. For example, A $(3,0,0)$ means
that Dimension~A is the dominant dimension, with 3 successful branches in
Dimension~A, 0 in Dimension~B, and 0 in Dimension~C.}
\label{tab:dim_prior_validation}
\footnotesize
\setlength{\tabcolsep}{4pt}
\renewcommand{\arraystretch}{1.05}
\begin{tabular}{lccc}
\toprule
\textbf{Failure category}
& \textbf{Pilot}
& \textbf{RDB Main Test}
& \textbf{WDJ-U} \\
\midrule
\texttt{wrong\_strategy}
& A $(3,0,0)$
& A $(4,1,0)$
& A $(20,0,1)$ \\
\texttt{wrong\_target}
& A $(2,0,0)$
& A $(5,0,1)$
& A $(7,1,0)$ \\
\texttt{env\_boundary}
& A $(2,1,0)$
& A $(4,1,0)$
& A $(9,2,0)$ \\
\bottomrule
\end{tabular}
\end{table}

\subsection{Branch Generation Prompt}
\label{sec:appendix_branch_prompt}

The branch generation prompt (used in \S\ref{sec:probe_gen}) instructs the executor agent to produce $N$ candidate plans across the three operator dimensions. The prompt provides:
\begin{itemize}[nosep,leftmargin=*]
\item The task description and failure analysis from the FDS.
\item The trajectory tail (last steps before failure) for context.
\item The current screenshot at the fork-point state.
\item Per-dimension guidance and minimum count constraints (e.g., ``Dimension~A minimum: 3, Dimension~B minimum: 1, Dimension~C minimum: 1'').
\end{itemize}

Each generated plan is a JSON object with fields: \texttt{dimension/type} (A/B/C), \texttt{title} (short identifier), \texttt{plan} (step-by-step concrete strategy), and \texttt{reason} (why this plan addresses the identified failure). The dimension guidance for Type~A is dynamically generated from the FDS (\texttt{restart\_explanation}); Type~B and C use fixed templates targeting viewport expansion and interactability probing respectively.

\paragraph{Trajectory Reuse and Incremental Graph Update.}
Each executed probe contributes a trajectory fragment consisting of the resumed context, the action sequence taken under the probe, the observed UI states, and the terminal outcome label. We append the probe type, the realized fragment, and the success/fail outcome to $\mathcal{H}_k$, and merge newly observed states and transitions into $\noisegraph_k^{\cup}$ by aliasing repeated screenshots / DOM configurations to previously seen local states whenever the UI context is unchanged. When resuming from an intervention point, the executor replays or re-establishes the prefix needed to recover the local UI state around $t_k^*$, then launches the new probe from that recovered context rather than discarding prior exploration and restarting the entire case.

\subsection{Information-Value Branch Ranking Prompt}
\label{sec:appendix_selection_prompt}

Branch ranking is conditioned on the current Failure Diagnostic Summary (FDS) $\mathcal{F}_k$ together with local execution context, including the visible UI state at $t_k^*$ and prior probe outcomes in $\mathcal{H}_k$, rather than unconstrained free-form text alone.
The branch selection module (\S\ref{sec:eig_ranking}) instructs the supervisor to rank and select the top-$K$ executable branches from the candidate pool by approximating the EIG ranking score in Eq.~\ref{eq:eig}. Each candidate branch is evaluated with respect to both attribution relevance and executability.
The prompt first applies four validity checks:
\begin{enumerate}[nosep,leftmargin=*]
\item \textbf{Root-cause fit}: does the branch directly address the diagnosed failure mode or dominant SOU?
\item \textbf{Feasibility}: can the branch be executed from the current checkpoint and visible interface state?
\item \textbf{Non-redundancy}: does the branch avoid trivially repeating actions that already failed without introducing new evidence?
\item \textbf{Specificity}: is the branch a concrete, executable action plan rather than a vague strategy?
\end{enumerate}

Branches that fail these checks are removed from $\mathcal{A}_{\mathrm{valid}}(\mathcal{S}_k)$. For each remaining branch, the supervisor estimates the branch outcome under the two latent attribution hypotheses: whether the branch is likely to succeed if the original failure was \textsc{AgentFail}, and whether the same outcome would remain likely if the environment is genuinely blocked. Operationally, the prompt asks for structured fields:
\begin{itemize}[nosep,leftmargin=*]
\item \texttt{valid}: whether the branch passes the feasibility, specificity, and non-repetition checks.
\item \texttt{p\_success\_agent}: estimated likelihood of success under \textsc{AgentFail}.
\item \texttt{p\_success\_env}: estimated likelihood of an observed success signal under \textsc{EnvFail}.
\item \texttt{eig\_rationale}: why the success/failure outcomes would or would not separate the two attribution hypotheses.
\item \texttt{rank}: the final order among valid branches by expected attribution-entropy reduction.
\end{itemize}

These numeric likelihoods are not treated as globally calibrated probabilities. Instead, they instantiate the local world model used for branch prioritization at selection time. Formally, for a branch $b$ of type $d(b)$, we use
\[
\begin{aligned}
w_b &= P(\texttt{verified\_success}\mid\textsc{AgentFail},b),\\
\beta_{d(b)} &= P(\texttt{verified\_success}\mid\textsc{EnvFail},b),\\
\gamma_{d(b)} &= P(\texttt{fail}\mid\textsc{AgentFail},b),
\end{aligned}
\]
where $w_b$ captures recovery under evaluator-side failure, $\beta_{d(b)}$ captures success that can still occur under environment-side failure, and $\gamma_{d(b)}$ captures failure compatibility under evaluator-side failure. The actual observed outcome is incorporated afterward through the branch-typed score update in Eq.~\ref{eq:update_rule}. Therefore, selection-time information value acts as a planning signal for branch ordering, while the realized entropy reduction is determined only after branch execution.

\section{Case Study and Examples}
\subsection{Diagnostic Pipeline Case Example}
\label{sec:appendix_pipeline_case}

We walk through a complete \textsc{DiagEval} diagnostic cycle on a concrete case from RealDevBench to illustrate how the components from \S\ref{sec:method} interact in practice.

\paragraph{Task.} ``Check saturation adjustment functionality'' on a film-effect image processing website.

\paragraph{Failed trajectory (Round 1).} The agent attempted to interact with a saturation slider before loading any image into the editor. Clicks at coordinates (960, 692) on the ``Choose File'' button did not trigger a file dialog. After repeated failed clicks and three action-error retries, the agent reported failure without trying alternative approaches. The trajectory tail:

\begin{quote}\small\ttfamily
Step 2: click(960, 692) --- Click ``Choose File'' to upload image\\
Step 3: click(1850, 442) --- Click saturation slider\\
Step 4: click(960, 692) --- Retry ``Choose File'' button\\
Steps 5--7: [action error, retrying] --- Report failure
\end{quote}

\paragraph{FDS output.}

\begin{quote}\small
\textbf{failure\_type}: agent \quad\textbf{failure\_category}: \texttt{wrong\_target}\\
\textbf{fail\_reason}: ``The agent attempted to interact with a saturation slider before any image was loaded into the editor, and the file upload button clicks at (960, 692) did not successfully open a file dialog or load an image. The agent never established a valid starting state (an image in the editor) before trying to test the saturation adjustment feature.''\\
\textbf{should\_retry}: true \quad\textbf{initial $p_{\text{env}}$}: 0.10\\
\textbf{fork-point ($t_0^*$)}: step 0 (complete restart)\\
\textbf{restart\_explanation}: ``No step produced verifiable progress, so restart from step~0. The agent should first take a screenshot to understand the current UI state, then locate and click the correct file upload button, confirm an image is loaded in the editor, and only then interact with the saturation slider.''
\end{quote}

The low initial $p_{\text{env}}=0.10$ reflects the supervisor's assessment that this is likely an agent-side grounding error rather than a broken application.

\paragraph{Dimension allocation.} Given \texttt{failure\_category = wrong\_target}, the empirical priors from Table~\ref{tab:dim_priors} assign weights $w_A{=}0.6$, $w_B{=}0.2$, $w_C{=}0.2$, concentrating the branch budget on alternative transitions (Type~A), because \texttt{wrong\_target} failures are most likely resolved by finding the correct interactive element.

\paragraph{Generated branches (5 candidates).}

\begin{enumerate}[nosep,leftmargin=2.4em,labelsep=0.45em]
\item[\textbf{A-0}:] \emph{OCR-based button targeting.} Use visual text recognition to find the ``Upload''/``Open''/``Choose File'' button instead of hardcoded coordinates; click the detected bounding box center.
\item[\textbf{A-1}:] \emph{Parent container interaction.} Click the larger drop-zone container surrounding the upload area with a 50px offset from the previously failed point.
\item[\textbf{A-2}:] \emph{Alternative image source.} Look for a ``Sample Image'' gallery or ``File'' menu in the navigation bar to load an image without using the upload button.
\item[\textbf{B-3}:] \emph{Viewport expansion.} Scroll down 500px to check if the upload button or saturation slider is partially off-screen.
\item[\textbf{C-4}:] \emph{Interactability probe.} Hover over (960, 692) to check cursor change, confirming whether the element is actually interactive before clicking.
\end{enumerate}

\paragraph{Selected branches ($K{=}3$).} The supervisor selected A-0, A-2, A-1 in priority order, all of which are Type~A, consistent with the \texttt{wrong\_target} SOU hypothesis. Under the local world model, these branches have high EIG score because success under any of them would strongly support \textsc{AgentFail}: the application would be shown reachable once the upload step is grounded correctly. Their failure would also provide useful negative evidence because three distinct action-form realizations would have failed from the same fork point. The B and C candidates were deprioritized because viewport expansion and hover-only interactability checks are less likely to separate \textsc{AgentFail} from \textsc{EnvFail} for this diagnosed grounding error.

This example illustrates the full diagnostic pipeline: the FDS produces a structured failure analysis with SOU hypothesis, branch generation proposes candidate probes over diagnostic dimensions, information-value selection prioritizes probes expected to maximally shift the attribution signal, and executed outcomes update the internal attribution score.
\vspace{-0.4em}
\paragraph{Intervention-point execution details.}
Operationally, resuming from $t_k^*$ means recovering the local UI context needed for the next diagnostic action, not teleporting to an exact simulator state. When available, the system replays the verified trajectory prefix up to $t_k^*$; otherwise it reconstructs the local context by navigating back to the same page / panel configuration before executing the probe. This design is sufficient for diagnosis because probes are defined over the local neighborhood of $t_k^*$, but it can be imperfect when the environment contains hidden session state or asynchronous effects that are difficult to reproduce exactly.

\subsection{Illustrative Running Example}
\label{sec:appendix_running_example}
\vspace{-0.4em}
\paragraph{Running example.}
Consider the test point \emph{``verify that clicking the timeline sets the
video end point''} in a VideoClipper application. The initial evaluation
fails because the file-upload button is non-responsive—the evaluator
cannot reach the timeline editor. \textsc{DiagEval} classifies this as
\texttt{env\_boundary} and allocates branch budget with emphasis on
Dimension~A.
Starting from the
maximum-ignorance prior $p^{(0)}_{\textsc{env}} = 0.50$, the supervisor
selects three probes spanning all three evidence channels:

\begin{enumerate}[leftmargin=1.5em, itemsep=2pt]
  \item \textbf{Branch~1 (Dim~B, observation probe):} scroll the page to
        search for a pre-loaded sample video or alternative entry point.
        \emph{Fails.}
        $p^{(1)}_{\textsc{env}} = 0.50
          \xrightarrow{\gamma_B=0.50} 0.67$
        \;($\Delta p = +0.17$).
  \item \textbf{Branch~2 (Dim~C, diagnostic probe):} Tab-cycle through
        focusable elements to test whether the upload button accepts
        keyboard activation.
        \emph{Fails.}
        $p^{(2)}_{\textsc{env}} = 0.67
          \xrightarrow{\gamma_C=0.40} 0.83$
        \;($\Delta p = +0.17$)—this probe uses the smallest $\gamma$ value,
        making failure hardest to reconcile with evaluator-side error and
        therefore providing the strongest evidence toward environment-side
        blockage.
  \item \textbf{Branch~3 (Dim~A, re-instantiation probe):} hover and
        double-click the upload button with alternative timing.
        \emph{Fails.}
        $p^{(3)}_{\textsc{env}} = 0.83
          \xrightarrow{\gamma_A=0.60} 0.89$
        \;($\Delta p = +0.06$)—the smallest shift, because
        $\gamma_A{=}0.60$ is the largest likelihood, making a failed
        re-instantiation most compatible with evaluator miss. The
        diminishing increment also reflects score saturation: as
        $p_{\textsc{env}}$ approaches~1, each additional failure
        contributes less marginal evidence.
\end{enumerate}

\begin{figure}[tbp]
    \centering
    \includegraphics[width=0.82\columnwidth]{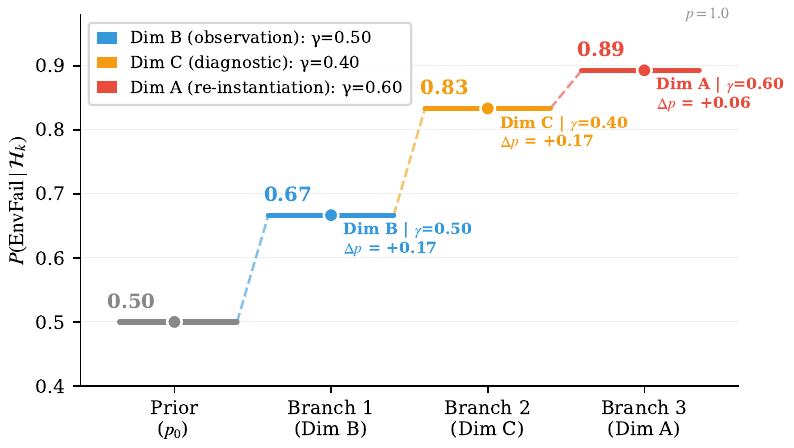}
    \caption{\textbf{Attribution-score update across diagnostic branches in the
    running example.} Each branch yields the same binary outcome (fail),
    but produces a different update magnitude due to branch-typed
    likelihoods $\gamma_d$, consistent with
    Eq.~\ref{eq:update_rule}.}
    \label{fig:posterior_staircase_appendix}
\end{figure}

\noindent
After exhausting the diagnostic budget, the internal attribution score
$p_{\textsc{env}} \approx 0.89$ strongly favors environment-side blockage
(Figure~\ref{fig:posterior_staircase_appendix}). All three branches
produce the same binary outcome (fail), yet the branch-typed observation
model converts them into different evidence strengths via the
dimension-specific likelihoods $\gamma_d$. This is the core mechanism that
enables structured attribution evidence: uniform retry, which treats every failure
identically, cannot distinguish these cases.

\subsection{Additional Case Studies}
\label{sec:appendix_cases}

\begin{casebox}{Case 3: Trajectory anchoring (OfficeEmployeeTaskAllocator)}
\textbf{Observed mismatch and resume-only behavior.} Baseline sees dashboard showing 6 tasks but ``All Tasks'' list shows only 4. NR+IE ($\times$1, $\times$2) resumes and repeats the same observation three times, anchored to ``count rows in All Tasks list.''\par
\textbf{\textsc{DiagEval} branch and outcome.} NR ($\times$2) discovers an alternative via status filters: Not Started~(1) + In Progress~(3) + Completed~(2) = 6.\par
\textbf{Takeaway.} Resume context without branching locked the agent into a failed verification strategy.
\end{casebox}

\begin{casebox}{Case 4: Multi-step architectural reasoning (Professional Portfolio)}
\textbf{Observed failure and resume-only behavior.} Baseline tries Download button, Ctrl+S, More actions menu, right-click, and no download is triggered. NR ($\times$2) repeats the same button-clicking pattern with no memory of what was tried.\par
\textbf{\textsc{DiagEval} branch and outcome.} \textsc{DiagEval} ($\times$2) identifies that direct button clicks are ineffective and generates a Dim~A branch: open ``Download Resume'' $\rightarrow$ PDF viewer loads $\rightarrow$ click download icon within viewer $\rightarrow$ save dialog triggered.\par
\textbf{Takeaway.} This two-step approach requires understanding the application's document viewing architecture.
\end{casebox}

\begin{casebox}{Case 5: Environment pre-conditioning (CSS Animation Playground)}
\textbf{Observed failure.} No method succeeds in $\times$1 because the animation completes before the pause button can be tested.\par
\textbf{\textsc{DiagEval} branch and outcome.} \textsc{DiagEval} ($\times$2) diagnoses the temporal root cause and generates a Dim~B branch: set ``Iteration Count'' to Infinite \emph{first}, then click pause. The animation stops and the button icon changes.\par
\textbf{Takeaway.} The environment is functional but requires a specific interaction ordering.
\end{casebox}

\section{Additional Experiments}

\subsection{Cross-Framework Transfer}
\label{app:transfer}
\vspace{-0.4em}
We further evaluate whether \textsc{DiagEval} can improve a different GUI-agent
framework without retuning. Specifically, \textsc{DiagEval} is developed with
AppEvalPilot as the source framework, and we transfer the same diagnostic
procedure to UI-TARS as the target framework. The diagnostic components,
including failure parsing, branch generation, branch selection, probing, and
attribution-score update, are kept unchanged; only the underlying GUI-agent
framework is replaced.

\begin{figure}[t]
    \centering
    \includegraphics[width=0.8\textwidth]{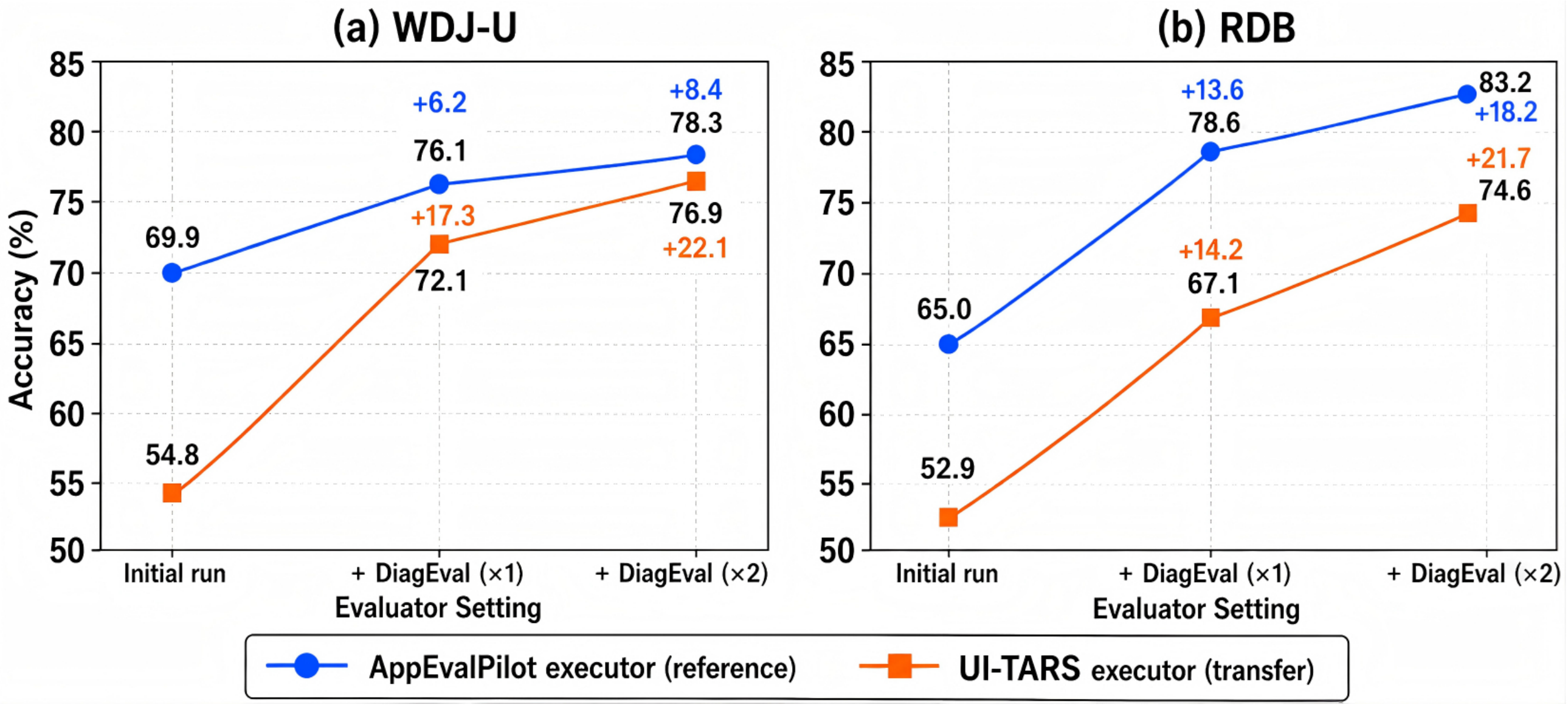}
    \captionof{figure}{\textbf{Cross-framework transfer results.}
    $\Delta$ denotes the absolute accuracy gain in percentage points over the
    corresponding single-pass baseline of the same GUI-agent framework.}
    \label{fig:transfer}
\vspace{-0.7em}
\end{figure}

Figure~\ref{fig:transfer} reports absolute accuracy changes in percentage
points over each framework's own single-pass baseline. On WDJ-U, UI-TARS obtains
54.8\% accuracy under single-pass evaluation. Adding \textsc{DiagEval} raises
accuracy to 72.1\% under the $\times1$ setting and 76.9\% under the $\times2$
setting, corresponding to absolute gains of $+17.3$ and $+22.1$ percentage
points. On RDB, UI-TARS improves from 52.9\% to 67.1\% and 74.6\%, giving
absolute gains of $+14.2$ and $+21.7$ percentage points. These results show that
\textsc{DiagEval} provides substantial absolute accuracy improvements beyond
the framework on which it was developed, indicating that the diagnostic
procedure transfers across GUI-agent frameworks rather than relying on
AppEvalPilot-specific behavior.

\vspace{0.75em}

To test whether the gains of \textsc{DiagEval} can be explained by repeated sampling alone, we compare it against a diagnosis-free \emph{naive retry} baseline on the same WDJ-U originally false-negative subset used in the main evaluation ($114$ cases). For a fair comparison, naive retry uses the same backbone (Gemini 3 Flash Preview) and the same AppEvalPilot framework as \textsc{DiagEval}, so that any difference reflects the diagnostic mechanism rather than backbone capacity. Naive retry reruns each task from scratch, without failure parsing, branch construction, or attribution-score updates.
We consider three naive-retry baselines: \emph{Retry-1} (a single retry), \emph{Majority-3} (majority vote over three attempts), and \emph{Best-of-3}, which counts a case as recovered if any attempt succeeds. Best-of-3 is therefore an optimistic sampling-only baseline.

\begin{table}[H]
\centering
\footnotesize
\setlength{\tabcolsep}{4pt}
\renewcommand{\arraystretch}{1.08}
\caption{\textbf{Recovery on WDJ-U originally false-negative cases.}
All methods are evaluated on the same 114 cases. Retry-1 denotes a single naive retry; Majority-3 denotes majority vote over three attempts; Best-of-3 counts a case as recovered if any attempt succeeds.}

\label{tab:appendix_naive_retry_wdju}
\begin{tabular}{lc}
\toprule
\textbf{Method} & \textbf{Recovered / Recovery Rate} \\
\midrule
Naive retry (Retry-1)    & $20/114$ \; $(17.5\%)$ \\
Naive retry (Majority-3) & $16/114$ \; $(14.0\%)$ \\
Naive retry (Best-of-3)  & $33/114$ \; $(28.9\%)$ \\
\midrule
\textsc{DiagEval} ($\times 1$) & $52/114$ \; $(45.6\%)$ \\
\textsc{DiagEval} ($\times 2$) & $68/114$ \; $(59.6\%)$ \\
\bottomrule
\end{tabular}
\end{table}

Table~\ref{tab:appendix_naive_retry_wdju} shows that \textsc{DiagEval}'s gains cannot be attributed to repeated sampling alone. Under naive retry, recovery remains limited on false negatives: one retry recovers $17.5\%$, Majority-3 recovers $14.0\%$, and Best-of-3 recovers $28.9\%$. The lower Majority-3 result also suggests that majority aggregation can introduce additional voting noise when retry outcomes are unstable across runs.
Even against the optimistic Best-of-3 baseline, \textsc{DiagEval} remains substantially stronger, reaching $45.6\%$ recovery after one diagnostic round and $59.6\%$ after two. This shows that \textsc{DiagEval} gains its advantage by identifying attribution-informative sources of uncertainty and converting them into targeted probes that help separate \textsc{AgentFail} from \textsc{EnvFail}, thereby reducing attribution uncertainty, rather than by retrying more times alone.

\vspace{-0.45em}
\subsection{Branch Prioritization}
\label{app:branch_prioritization_results}

We evaluate whether belief-conditioned branch prioritization improves diagnostic
efficiency. The analysis is conducted on the WebDevJudge-Unit FN set. Among the
114 FN cases, 50 are successfully recovered by diagnostic probing, and these
50 recovered cases form the evaluation set for branch ordering. This evaluation
is conditional on branch selection: for each case, the LLM-selected branch set
already contains at least one branch that can recover the task. We therefore keep
the selected branch set fixed and vary only the execution order, so that any
performance difference reflects the efficiency of ordering rather than the
ability to select a successful branch.

\vspace{-0.25em}
\par\noindent\textbf{Compared methods.}
We compare three ordering strategies:
(i) \textbf{Random ordering}, which executes the selected branches in random
order;
(ii) \textbf{LLM-default ordering}, the order produced by the LLM judge and used
as the default execution order in \textsc{DiagEval};
(iii) \textbf{EIG ordering}, our belief-conditioned ordering strategy that ranks
branches by EIG score over the attribution score.

\begin{table}[H]
\centering
\caption{\textbf{Ordering performance on recoverable cases.}
Different execution orders are compared with the selected branch set fixed.
\texttt{First-branch success} denotes the fraction of cases recovered by the first executed branch.}
\label{tab:eig_recovered_cases}
\footnotesize
\setlength{\tabcolsep}{4pt}
\renewcommand{\arraystretch}{1.08}
\begin{tabular}{lcccc}
\toprule
\textbf{Ordering strategy}
& \textbf{First-branch success}
& \textbf{Avg. probes / case}
& \textbf{Total probes}
& \textbf{Saving vs. random} \\
\midrule
Random ordering
& 33\% (17/50)
& 2.00
& 100
& -- \\
LLM-default ordering
& 58\% (29/50)
& 1.60
& 80
& 20\% \\
\textbf{EIG ordering}
& \textbf{66\% (33/50)}
& \textbf{1.48}
& \textbf{74}
& \textbf{26\%} \\
\bottomrule
\end{tabular}
\end{table}

As shown in Table~\ref{tab:eig_recovered_cases}, EIG ordering achieves the highest first-branch success rate and the lowest probe
cost. Relative to LLM-default ordering, it improves first-branch success
from 58\% to 66\% and reduces the total number of probes from 80 to 74. Compared
with random ordering, EIG doubles the first-branch success rate from 33\% to
66\% and saves 26\% of probe executions. These results suggest that, once the LLM judge has selected semantically plausible diagnostic branches, EIG-based ordering can more effectively place recoverable branches earlier in the execution sequence.

\subsection{Sensitivity to Hyperparameters}
\label{app:ablation_sensitivity}
\vspace{-0.25em}
\paragraph{Parameter-setting protocol.}
The default update and stopping parameters are empirical scoring constants rather than fitted probabilities.
We select $(w_0,\beta_0)=(0.60,0.20)$, $(\gamma_A,\gamma_B,\gamma_C)=(0.60,0.50,0.40)$, and $\tau_{\mathrm{env}}=0.70$ through lightweight sanity checks on the 41-case RDB pilot set, while holding out the 429-case RDB main test set for final reporting.
Table~\ref{tab:dim_priors} provides the pilot-derived routing sanity check for branch-dimension allocation, whereas Tables~\ref{tab:likelihood_sensitivity} and~\ref{tab:tau_threshold_sensitivity} test whether the reported trends are stable around the default scoring constants.

\par\noindent\textbf{Sensitivity to Likelihood Parameters}
\label{app:likelihood_sensitivity}
We test whether the attribution signal depends strongly on a particular choice of likelihood parameters (Table~\ref{tab:likelihood_sensitivity}). To align with the main method, we parameterize the branch-typed update by a base pair $(w_0,\beta_0)$, where $w_0$ denotes the nominal probability of observing \texttt{verified\_success} under \textsc{AgentFail}, and $\beta_0$ denotes the nominal probability of observing success under \textsc{EnvFail}. Branch-specific terms $w_b$ and $\beta_{d(b)}$ are instantiated from this base pair together with branch type.

Across settings, \textsc{DiagEval} remains stable and retains non-trivial entropy reduction. Even at $w_0=\beta_0=0.50$, where success/failure outcomes alone carry no attribution preference, \textsc{DiagEval} still achieves $\Delta H \approx 0.25$--$0.27$. This indicates that attribution signal does not come only from raw success/failure outcomes, but also from the branch-typed probing structure itself. Increasing the separation between $w_0$ and $\beta_0$ yields larger $\Delta H$, but also makes the update more sensitive to the assumed likelihood asymmetry. We therefore use $(w_0,\beta_0)=(0.60,0.20)$ as the default setting in the main experiments.

\begin{table}[H]
\centering
\caption{\textbf{Sensitivity to likelihood parameters for \textsc{DiagEval} ($\times 1$).}
Results are reported in terms of $\Delta H$ under the branch-typed score update. The row $w_0=\beta_0=0.50$ is a \emph{non-informative reference}, in which success/failure outcomes alone provide no directional preference between \textsc{AgentFail} and \textsc{EnvFail}; any remaining attribution signal therefore comes from the branch-typed probing structure. The default setting $(w_0,\beta_0)=(0.60,0.20)$ is used in the main experiments.}
\footnotesize\setlength{\tabcolsep}{3pt}\renewcommand{\arraystretch}{0.95}
\begin{tabular}{@{}lcc@{}}
\toprule
$w_0$ / $\beta_0$ & \textbf{RDB} & \textbf{WDJ-U} \\
\midrule
$w_0{=}\beta_0{=}0.50$ (non-inform.\ ref.) & 0.269 & 0.254 \\
$w_0{=}0.50$, $\beta_0{=}0.10$ & 0.365 & 0.361 \\
$w_0{=}0.60$, $\beta_0{=}0.20$ (default) & \textbf{0.305} & \textbf{0.274} \\
$w_0{=}0.70$, $\beta_0{=}0.20$ & 0.321 & 0.315 \\
$w_0{=}0.80$, $\beta_0{=}0.10$ & 0.428 & 0.429 \\
$w_0{=}0.90$, $\beta_0{=}0.05$ & 0.531 & 0.536 \\
\bottomrule
\end{tabular}
\label{tab:likelihood_sensitivity}
\end{table}

\vspace{-0.25em}
\par\noindent\textbf{Stopping Threshold Sensitivity}
\label{app:stoppint_params}
We also vary the \textsc{EnvFail} stopping threshold $\tau_{\mathrm{env}}$
(Table~\ref{tab:tau_threshold_sensitivity}), which controls when diagnosis
terminates early and returns an \textsc{EnvFail} verdict. The stopping threshold
$\tau_{\mathrm{env}}$ is designed to affect efficiency: it determines how early
diagnosis terminates once the accumulated \textsc{EnvFail} belief becomes
sufficiently high. In our results, $\tau_{\mathrm{env}}=0.70$ provides the best
trade-off: it saves $67$ branches ($35.4\%$) while preserving perfect
recoverable-case recovery ($50/50$) and incurring zero truncation error.

\begin{table}[H]
\centering
\caption{\textbf{Sensitivity of the \textsc{EnvFail} stopping threshold $\tau_{\mathrm{env}}$.}
We vary the early-stopping threshold $\tau_{\mathrm{env}}$ for the accumulated
\textsc{EnvFail} belief and report its effect on diagnostic efficiency and
recoverable-case safety. \texttt{Saved branches} denotes the number and
percentage of diagnostic branches skipped by threshold-based early stopping.
\texttt{Correct RC recovery} counts recoverable cases that are still successfully
recovered, while \texttt{RC truncation error} counts recoverable cases that are
incorrectly stopped before recovery.}
\label{tab:tau_threshold_sensitivity}
\small
\setlength{\tabcolsep}{7pt}
\renewcommand{\arraystretch}{1.12}
\begin{tabular}{cccc}
\toprule
\textbf{Threshold $\tau_{\mathrm{env}}$}
& \textbf{Saved branches}
& \textbf{Correct RC recovery}
& \textbf{RC truncation error} \\
\midrule
0.65 & 83 (43.9\%) & 37/50 & 13/50 \\
\textbf{0.70} & \textbf{67 (35.4\%)} & \textbf{50/50} & \textbf{0/50} \\
0.75 & 40 (21.2\%) & 50/50 & 0/50 \\
0.80 & 20 (10.6\%) & 50/50 & 0/50 \\
0.85 & 1 (0.5\%) & 50/50 & 0/50 \\
0.90 & 0 (0.0\%) & 50/50 & 0/50 \\
\bottomrule
\end{tabular}
\end{table}

The threshold results show that the efficiency gain from early stopping must be
balanced against recoverable-case safety. A low threshold such as
$\tau_{\mathrm{env}}=0.65$ saves the most branches, but it prematurely stops 13
recoverable cases and reduces correct recovery to 37/50. In contrast, thresholds
of 0.75 and above preserve all recoverable cases, but the saved branches decrease
substantially as the threshold becomes more conservative. The setting
$\tau_{\mathrm{env}}=0.70$ achieves the most favorable operating point in this
experiment: it retains perfect recovery on recoverable cases while still skipping
over one third of diagnostic branches. Together with the ordering results, this
suggests that belief-conditioned diagnosis improves efficiency in two
complementary ways: EIG ordering reduces the number of probes needed to find a
successful branch, while threshold-based early stopping avoids unnecessary
probing once the evidence for \textsc{EnvFail} is sufficiently strong.

\subsection{Model-Call Composition in FN Retry}
\label{app:model_call_composition}

\begin{figure}[H]
    \centering
    \includegraphics[width=0.78\textwidth]{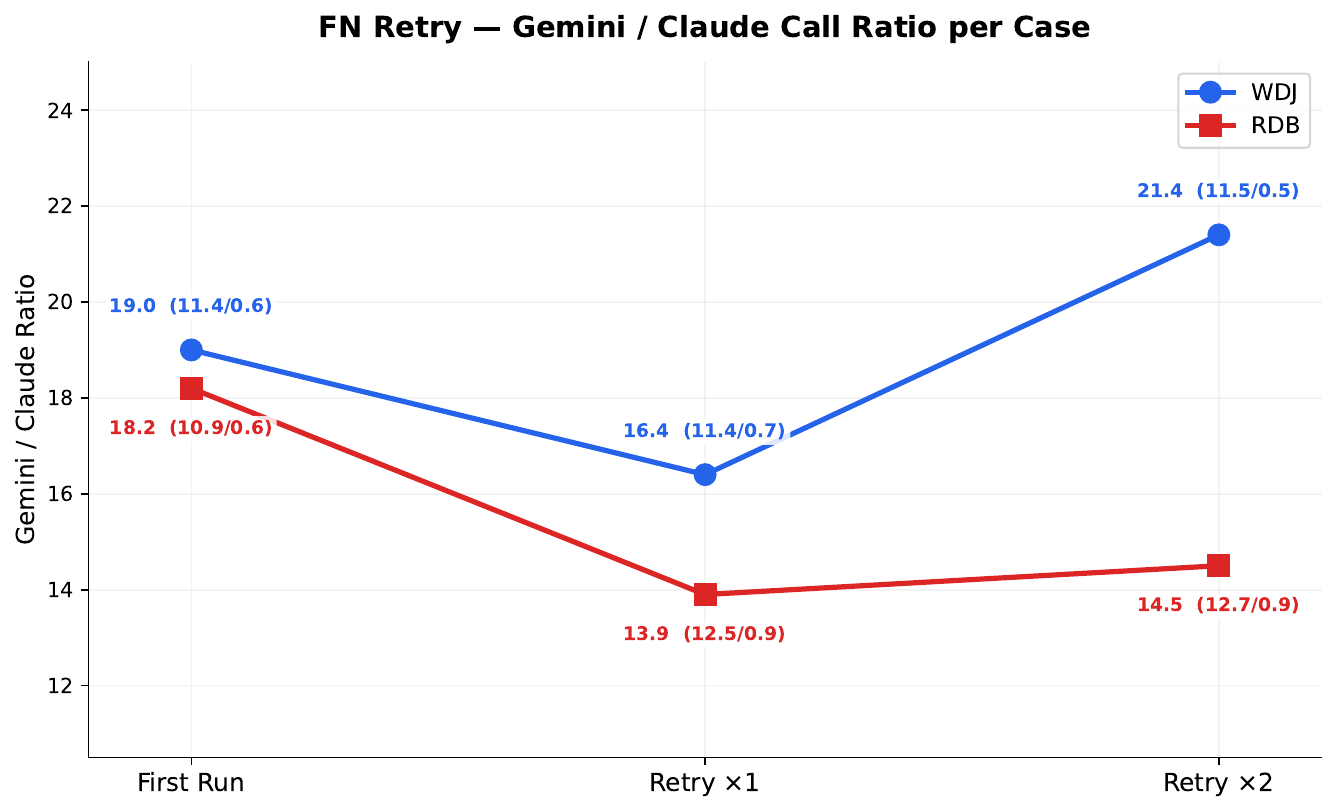}
    \caption{\textbf{Gemini / Claude call ratio per case across the FN retry pipeline.}
    The ratio is computed as the average number of Gemini agent calls divided by
    the average number of Claude supervisor calls per case. Parenthetical values
    report the rounded underlying averages (Gemini avg. / Claude avg.). Gemini
    calls remain relatively stable across stages, while most variation in the
    ratio comes from Claude-side diagnostic supervision.}
    \label{fig:fn_retry_model_calls}
\end{figure}

Figure~\ref{fig:fn_retry_model_calls} reports the Gemini/Claude call ratio per
case across the FN retry pipeline. This ratio reflects the division of labor in
\textsc{DiagEval}: Gemini is used for the GUI-agent execution loop, while Claude
is used for supervisor-side diagnosis, retry-plan generation, and checkpoint
analysis. In the first run, both benchmarks show a high ratio
($\approx$18--19), since each case mainly invokes Gemini for the agent rollout
and only requires lightweight Claude-side supervision. After Retry~$\times1$,
the ratio decreases on both WDJ-U and RDB, reflecting additional Claude calls
introduced by failure verification, diagnostic planning, and restart-point
analysis.

Across all stages, the average number of Gemini calls remains within a relatively
narrow range ($\approx$11--13 per case), suggesting that the agent-side execution
cost is largely determined by the fixed rollout budget. In contrast, the
Gemini/Claude ratio varies mainly with Claude-side supervision. By
Retry~$\times2$, WDJ-U shows a higher ratio because its average Claude-call count
decreases, whereas RDB remains lower due to more sustained supervisor
involvement. Thus, this figure complements the cost analysis by showing that
variation in per-case call composition is driven primarily by diagnostic
supervision rather than by large changes in the Gemini execution loop.


\end{document}